\documentclass[aps,prd,twocolumn,amsmath,amssymb,superscriptaddress,nofootinbib,showpacs,showkeys]{revtex4-1}

\usepackage{graphicx}
\usepackage{hyperref}    
\begin{document}

\def\t{\tilde}

\title{Combined cosmological and solar system constraints on chameleon mechanism}

\author{A.~Hees}
\email{aurelien.hees@oma.be}
\affiliation{Royal Observatory of Belgium \\ Avenue Circulaire 3, 1180 Bruxelles, Belgium}
\affiliation{LNE-SYRTE, Observatoire de Paris}
\affiliation{Namur Center for Complex systems (naXys),\\ University of Namur, Belgium}   

\author{A.~F\"uzfa}
\email{andre.fuzfa@fundp.ac.be}
\affiliation{Namur Center for Complex systems (naXys),\\ University of Namur, Belgium}
\affiliation{Center for Cosmology, Particle Physics and Phenomenology (CP3), \\ University of Louvain, Belgium}

\date{November, 2011}


\begin{abstract}
Chameleon mechanism appearing in massive tensor-scalar theory of gravity can effectively reduce locally the non-minimal coupling between the scalar field and matter. This mechanism is invoked to reconcile large-scale departures from general relativity (GR), supposedly accounting for cosmic acceleration, to small scales stringent constraints on GR. In this paper, we carefully investigate this framework on cosmological and solar system scales to derive combined constraints on model parameters, notably by performing  a non-ambiguous derivation of observables like luminosity distance and local post-newtonian parameters. Likelihood analysis of type Ia SuperNovae (SN Ia) data and of admissible domain for the PPN parameters clearly demonstrates that chameleon mechanism cannot occur in the same region of parameters space than the one necessary to account for cosmic acceleration with the assumed Ratra-Peebles potential and exponential coupling function. 
\end{abstract}

\maketitle

\section{Introduction}
Today, gravitation is facing a major problem: on one hand, General Relativity (GR) have passed all solar system experiments ; on the other hand, GR and the Standard  Model of particles are not sufficient to explain galactic and cosmological observations, for instance the accelerated expansion of the universe. Several options have been advanced to solve this apparent inconsistency. The most widespread possibility consists of extending the matter-energy content of the universe with new ingredients like dark matter and dark energy. Another possibility consists of modifying GR on large scales without the introduction of a new type of matter. The construction of a new theory of gravity is particularly hard because it has to satisfy very stringent constraints on solar system scales (see Will~\cite{will:1993fk,*will:2006cq} for a review) while showing deviations on large scales.
\\

Moreover from a theoretical point of view, the different attempts to quantize gravity or to unify it with other fundamental interactions predict deviations from GR. In most of the theories proposed to account for cosmic acceleration, to quantize gravity or to unify it with the standard model,  scalar fields naturally appear in addition to the metric tensor. For examples, scalar fields appear in Kaluza-Klein theories or in string theories. They also appear in non-minimal extension of GR like in $f(R)$ theories~\cite{de-felice:2010uq} or in credible cosmological scenarios (inflation or quintessence).
\\

Tensor-scalar theories are therefore very important and they are  already widely studied in the literature. The first tensor-scalar theory was introduced by Jordan, Brans and Dicke~\cite{jordan:1949vn,brans:1961fk} in order to recover the Mach principle. A very detailed study of tensor-scalar theory can be found in Damour and Esposito-Far\`ese~\cite{damour:1992ys}. Tensor-scalar theories incorporate both the direct coupling of the scalar field to matter 
(through the coupling function $A(\phi)$ and a scalar self-interaction term (represented by a potential $V(\phi)$). 
Massless tensor-scalar theories (characterized by a vanishing potential $V(\phi)=0$) are strongly constrained at the linear level 
of the coupling function by solar system experiments~\cite{will:2006cq} and by binary pulsars~\cite{damour:1996uq}, yet rather poorly at the non-linear level. In particular, the observation of the Shapiro delay with Cassini spacecraft during a solar conjunction~\cite{bertotti:2003uq} gives currently the best constraint on the post-newtonian parameter $\t\gamma$
\begin{equation}   \label{cassIntro}
	\t\gamma-1=2.1\pm 2.3\times 10^{-5}\cdot
\end{equation}                              
This constraint implies that the coupling function at the linear level $d\log(A)/d\phi=k(\phi)\ll 1$~\footnote{In standard literature about tensor-scalar theory, the coupling constant is often noted $\alpha$ which we do not use in order to avoid confusions with the $\alpha$ parameters entering the Ratra-Peebles potential (\ref{potratra}).}. This low limit puts a severe constraint on some theoretical development that requires a coupling of the order of the unity (for example string theory or $f(R)$ gravity produce coupling of the order of the unity) unless some mechanism is reducing this coupling constant.
\\

In 2004, Khoury and Weltman~\cite{khoury:2004fk,khoury:2004uq} showed that certain massive tensor-scalar theory can satisfy solar system experiments even for high coupling constants. This happens because the amplitude of the mass of the scalar field depends on the local matter density. In particular, in a region of high density (for example in the Sun) the scalar field acquires a high mass while on cosmological scales (where the local density is low) the mass of the scalar field is very low. The fact that the scalar field has a  mass  mimicking the local density justifies the name given to these scalar field called \textit{chameleon fields}.                              
\\

Considering a spherical body, Khoury and Weltman~\cite{khoury:2004fk,khoury:2004uq} showed that under certain conditions, the scalar field is nearly frozen inside the body at the exception of a thin shell around the surface's body. This screening effect appears only if the so-called thin-shell parameter (noted $\varepsilon$) is small enough. This effect has the interesting property to reduce the effective scalar charge of the body that is equal to the coupling constant $k$ in usual tensor-scalar theory in low gravitational field (non perturbative effects can appear in strong field as shown in Damour and Esposito-Far\`ese~\cite{damour:1993vn}). Since the effective scalar charge entering in the expression of the $\t\gamma$ post-newtonian parameters is reduced, this mechanism allows the theory to pass the Cassini constraint (and more generally solar system experiments) even for high coupling constants $k$ that were previously excluded by PPN constraints.
\\

The main constraint on chameleon model on solar system scales is that the mass of the scalar field has to be sufficiently large to satisfy solar system constraints (equivalence principle, fifth force and PPN constraints). In Khoury and Weltman~\cite{khoury:2004fk}, it is explicitly argued that with this mechanism it is possible to explain cosmological effects without producing observable deviations in solar system because the environment is dense enough and the contribution from the scalar field is exponentially suppressed.
\\

The goal of this paper is to verify this assumption and to derive acceptable domain for the parameters characterizing the theory (mainly the coupling constant $k$ and the parameters characterizing the potential $V(\phi)$). The chameleon mechanism has already been widely studied in the literature~\cite{khoury:2004fk,khoury:2004uq,brax:2004fk,mota:2006kx,*mota:2007fk,faulkner:2007kx,tamaki:2008ve,tsujikawa:2009qf}. Nevertheless some issues need still to be addressed. First of all, none of these articles derive constraints from Supernovae data taking correctly  into account non minimal coupling. In this communication, we will use Supernovae data to find confidence region for the parameters characterizing the theory. Moreover, all publications related to chameleon fields used quantities in Einstein frame to perform calculations and to interpret results. In this communication, we review the mechanism by deriving unambiguously physical observables from Einstein and Jordan frames. The latter is defined as the set of gravitational degrees of freedom in which matter couples directly to a metric, although that metric does not propagate like it does in general relativity.  This distinction between Jordan and Einstein frame quantities has implications for cosmology because we used a Jordan frame conserved densities contrary to what is done in~\cite{brax:2004fk,mota:2007fk}. In the same spirit, we take a great care in the derivation of the observable (the luminosity distance). To avoid any confusion, we derive observables in both frames showing their physical equivalence.  

The same approach is applied for computing the static and spherical solution representing the solar system. Indeed, we use Jordan frame quantity in the field equations contrary to what was done in~\cite{khoury:2004fk,tamaki:2008ve,tsujikawa:2009qf} (but following the work of~\cite{babichev:2010fk}). On solar system scales, analytical solutions representing the scalar field were computed in \cite{khoury:2004fk,khoury:2004uq,mota:2006kx,*mota:2007fk,tamaki:2008ve,tsujikawa:2009qf} assuming a lot of hypothesis that have never been checked. In this paper, we compare these solutions with a numerical integration of the Einstein equations (in the same way as in~\cite{babichev:2010fk}). With these solutions, we derive the post-newtonian parameters taking care of the gauge used and we compare our results with the literature. Finally, we present a combined analysis of the cosmological study and the post-newtonian constraints.
\\

The strategy followed is to fit the model on real data at two different scales: on cosmological scales and on solar system scales. In this paper, we decide to focus only on the original theory proposed by Khoury and Weltman~\cite{khoury:2004fk,khoury:2004uq} which is a tensor-scalar theory with an exponential coupling function $A(\phi)=e^{k\phi}$ and a runaway potential $V(\phi)$ of the Ratra-Peebles type. This model is fully described in Section~\ref{sec:model}. We let for future work the study of other coupling constants and/or potentials.
\\

Concerning cosmological scales presented in Section~\ref{sec:cosmo}, we perform a likelihood analysis of the latest Supernovae Ia data. For this analysis, the cosmological evolution equations are derived and the expression of the luminosity distance is given. In the derivation of these equations, the use of the Einstein or Jordan frame quantities is fully discussed. In particular, the derivation of the luminosity distance has been performed in both frames showing their physical equivalence. From the likelihood analysis, we find confidence regions for the different parameters involved in the model. For models within these confidence regions, we also perform a full cosmological analysis consisting in a detailed study of the evolution of the scalar field $\phi$, the evolution of the cosmic expansion and its acceleration and we derive an effective dark-energy equation of state from the Friedmann-Lemaitre equations. With this analysis, we can clearly identify models explaining the accelerated expansion of the universe.
\\

In Section~\ref{ref:solarsystem}, we focus on solar system constraints. Solar system constraints are mainly composed of three tests: violation of the weak equivalence principle, fifth force constraints and Post-Newtonian constraints. In  our case, the weak equivalence principle is not violated since the coupling constant between the scalar field and matter is universal. However, we emphasize the fact that string theory can give raise to tensor-scalar theory with different coupling constants for the different matter fields. In this case, violation of the equivalence principle is expected~\cite{damour:2002vn}.
\\
In the model considered here a fifth force term is present. A fifth force analysis can be found in Khoury and Weltman~\cite{khoury:2004fk} while we concentrate our attention on the analysis of the post-newtonian parameters. To derive the expression of the post-newtonian parameters, we use the analytical solutions of the field profile in a static spherical configuration that can be found in the literature~\cite{khoury:2004fk,waterhouse:2006kx,tamaki:2008ve,tsujikawa:2009qf}. First of all, in order to check the consistency of the analytical expressions, we compare these analytical solutions (obtained by assuming a lot of hypothesis) with a full numerical determination of the field profile solving the Einstein equations representing a spherical and static body. Then, a careful derivation of the post-newtonian parameters is performed. With the models within the confidence region of the cosmological analysis, we compute the post-newtonian parameters $\t\gamma$ and we check if they satisfy the constraint (\ref{cassIntro}).            
\\

With these two analysis (cosmological and solar system analysis), it is possible to answer to the question: can the chameleon effect reconcile cosmological observations with solar system constraints ? For the model
treated here, the answer to this question is negative. Indeed, we will show that on one hand large coupling constants can passed the solar system constraints while they are excluded by usual analysis. On the other hand, tensor-scalar theories can indeed explain cosmic acceleration. But the main conclusion of this communication is that unfortunately, the parameters confidence region needed to explain cosmic acceleration does not overlap the one coming from solar system constraints. This means it is not possible to explain cosmic acceleration with the same parameters as the one passing solar system experiments.

\section{The model\label{sec:model}}  
The model studied in this paper is the one proposed in the original chameleon paper~\cite{khoury:2004uq,khoury:2004fk} and is given by the following action in the so-called Einstein frame ($c=1$):
\begin{eqnarray}
	S&=&\int d^4x\sqrt{-g}\left[\frac{m_p^2}{16\pi}R-\frac{m_p^2}{2}\partial_\mu\phi\partial^\mu\phi-V(\phi)\right] \nonumber\\
	&&+S_m\left[\Psi_m,A^2(\phi) g_{\mu\nu}\right]     \label{actionEinstein}
\end{eqnarray}                                                                                                                    
where $m_p=G^{-1/2}$ is the Planck mass, $g$ is the determinant of the Einstein-frame metric $g_{\mu\nu}$, R is the Ricci scalar and $\Psi_m$ are matter fields. In this frame, the scalar field $\phi$ interacts directly with matter particles through a conformal coupling characterized by the function $A(\phi)$. Finally $V(\phi)$ is a potential characterizing the scalar self-interaction. This action generalizes the Jordan-Fierz-Brans-Dicke theory (recovered in the case of a vanishing potential and of an exponential coupling function $A(\phi)=e^{k\phi}$)~\cite{jordan:1949vn,brans:1961fk} but also the quintessence model~\cite{wetterich:1988ys,ratra:1988vn} which can be recovered in the case $A(\phi)=1$. In this communication, we consider that all matter fields couple in the same way to the scalar field with only one coupling function $A(\phi)$. The above action can be justified by low-energy limit of string theory or supergravity~\cite{damour:1994fk,damour:1994uq,gasperini:2002kx,damour:2002ys,damour:2002vn}.

Observable quantities are not directly obtained in this frame because physical measurements are performed using rods and clocks built upon matter fields $\Psi_m$. These matter fields are universally coupled to the so-called Jordan frame metric 
\begin{equation}
	\tilde g_{\mu\nu}=A^2(\phi)g_{\mu\nu}
\end{equation}                           
and therefore follows geodesics corresponding to $\tilde g_{\mu\nu}$ and not to the Einstein frame metric $g_{\mu\nu}$. One can perform the conformal transformation in the action (\ref{actionEinstein}) to obtain the Jordan frame action (see for example~\cite{damour:1992ys,damour:1996uq,damour:1993kx,fuzfa:2009fk}). Nevertheless the Jordan frame is not convenient for studying the dynamics of the scalar field and of the metric $\tilde g_{\mu\nu}$ because the tensor and scalar modes are kinematically coupled and because this frame introduces some fictitious singularity~\cite{damour:1993kx,fuzfa:2009fk}.
                            
In the following, a $\tilde{}$ will denote quantities expressed in the Jordan frame. Observable quantities can easily be computed in this frame in the same way as in General Relativity (GR). This does not mean that observable quantities can only be derived in the Jordan frame. Indeed, physical predictions are not dependent of a change of variables in terms of a conformal transformation. If the derivation of observable quantities in Jordan frame is easy, one has to be more careful when working with the Einstein frame. In particular, all physical units have to be scaled with the coupling function $A(\phi)$.

The potential $V(\phi)$ in the action (\ref{actionEinstein}) is a key element in the chameleon theory. It has to be of the runaway form~\cite{khoury:2004uq,khoury:2004fk}. The potential used here is the Ratra-Peebles (RP) inverse law potential~\cite{wetterich:1988ys,ratra:1988vn} 
\begin{equation}     \label{potratra}
	V(\phi)=\frac{\Lambda^{\alpha+4}}{m_p^\alpha\phi^\alpha}.
\end{equation}
where $\alpha$ and $\Lambda$ are free parameters.

In this communication, we only consider an exponential coupling function
\begin{equation}
	A(\phi)=e^{k\phi}.
\end{equation}
This coupling function is the one used in the original paper~\cite{khoury:2004uq,khoury:2004fk} and we let the study of other type of coupling function to future work.

From variational principles, we can derive field equations from action (\ref{actionEinstein}). Varying this action with respect to the Einstein frame metric gives	 the Einstein field equations
\begin{eqnarray} \label{einstein}
  R_{\mu\nu}-\frac{1}{2}g_{\mu\nu}R&=&\frac{8\pi}{m_p^2}T_{\mu\nu}+8\pi\partial_\mu\phi\partial_\nu\phi\nonumber\\
&& -4\pi g_{\mu\nu}\partial_\alpha\phi\partial^\alpha\phi-\frac{8\pi}{m^2_p}g_{\mu\nu}V(\phi)
\end{eqnarray}
where $T_{\mu\nu}=-(2/\sqrt{-g})(\partial \mathcal L\sqrt{-g}/\partial g^{\mu\nu})$ is the Einstein frame stress-energy tensor.  

The variation of action (\ref{actionEinstein}) with respect to the scalar field $\phi$ leads to the Klein-Gordon equation for the scalar field:
\begin{equation}   \label{kleingordon}
	\Box \phi=-\frac{k(\phi)}{m^2_p}T+\frac{1}{m_p^2}\frac{dV}{d\phi}
\end{equation}                                                           
where $T=g^{\mu\nu}T_{\mu\nu}$ is the trace of the  stress-energy tensor, $\Box \phi=g^{\mu\nu}\nabla_\mu\nabla_\nu\phi$ and $k(\phi)$ is the scalar coupling strength to matter and is given by
\begin{equation}
	k(\phi)=\frac{d \ln A(\phi)}{d\phi}=\frac{1}{A(\phi)}\frac{d A(\phi)}{d \phi}.
\end{equation} 

Finally, the invariance of action (\ref{actionEinstein}) under coordinates transformations gives the following conservation equation in the Einstein frame:
\begin{equation}\label{consT}
	\nabla_\mu T^{\mu\nu}=k(\phi)T\partial^\nu\phi.
\end{equation}

In the following, we will always consider matter to be described by a perfect fluid. Therefore, the Einstein frame stress-energy tensor is expressed as
\begin{equation}
	T^{\mu\nu}=(\rho+p)u^\mu u^\nu+p g^{\mu\nu}
\end{equation}
where $\rho$ and $p$ are the Einstein frame density and pressure and $u^\mu$ is the 4-velocity of the fluid. As can be seen from (\ref{consT}), this stress-energy tensor is not conserved in the Einstein frame because of the explicit coupling between matter and the scalar field. On the other hand, the Jordan frame stress-energy tensor is conserved because matter field is universally coupled to the metric $\tilde g_{\mu\nu}$. Moreover, Jordan frame is defined as the frame in which matter experiences locally the laws of special relativity. The Jordan frame matter density and pressure ($\t\rho$ and $\t p$) are directly observable. The observable stress-energy tensor (Jordan frame) is related to its Einstein frame counterpart by a conformal factor~\cite{damour:1992ys,damour:1993kx}
\begin{eqnarray}  \label{stressenergy}
 T^{\mu}_{\quad\nu}&=&  A^4(\phi) \t T^{\mu}_{\quad\nu}\nonumber\\
 &=& A^4(\phi) (\t\rho+\t p)\t u^\mu \t u_\nu+ A^4(\phi) \t p \delta^\mu_\nu\nonumber\\
&=& A^4(\phi) (\t\rho+\t p) u^\mu  u_\nu+ A^4(\phi) \t p  \delta^\mu_\nu  .\nonumber
\end{eqnarray}
This implies the following relations between matter density and pressure in Einstein and Jordan frame:
\begin{subequations}\label{rhop}
\begin{eqnarray}
	   \rho  & =  &A^4(\phi)\t\rho \label{rho} \\
   p & =  &A^4(\phi)\t p.               \label{p}
\end{eqnarray}
\end{subequations}

In the following sections, we will derive observational constraints on different parameters characterizing the theory (namely parameters involved in the potential and the coupling function) and characterizing the cosmological content of the universe. 
First, we will perform a least-square fit on the SuperNovae Ia  data. This involves the derivation ant the integration of the equations describing the cosmological evolution of the universe. These equations will be derived from field equations (\ref{einstein}) and (\ref{kleingordon}).
Secondly, we will use solar system observations to obtain a different constraint on the theory parameters. To do this, we will derive the static spherical solution to the field equations. This solution will represent the solar system space-time. 

\section{Cosmological constraints\label{sec:cosmo}} 
This section is devoted to the study of the cosmological dynamics of the model described in the previous section. Let us briefly outline our original contributions concerning chameleon cosmology. We derive the evolution equation in Einstein frame but using density and pressure defined in Jordan frame. Therefore, we introduce a new variable $\tilde \Omega$ representing the observable density parameter.    In Sec.~\ref{sec:lum}, we derive a luminosity distance relation which is frame-independent (as any observables should be) and therefore allows us to make cosmological predictions without ambiguities. In particular, we present a likelihood analysis of the latest SNe Ia data's. These analysis extends previous works on quintessence models (characterized by a vanishing coupling constant $k=0$) to the cases of $k\ne 0$ where no analysis exist to our knowledge. We identify models statistically consistent with data and show show that the statistical adequacy decreases with $k$. The rest of the section comprises a detailed analysis of the model dynamics within the confidence regions previously established.  First of all, we extend the work of Brax et al~\cite{brax:2004fk} concerning the study of the dynamics of the scalar field by introducing three time-scales and we showed how the dynamics of $\phi$ depends on the parameters of the model. In Sec.~\ref{sec:cosmic}, we propose an original analysis of the cosmic expansion within the models explaining data. In this analysis, contributions due to the quintessence potential and to non-minimal coupling are clearly identified and compared. Finally, we interpret considered models in standard FLRW cosmology in General Relativity with an effective fluid whose quintessence and non-minimal coupling contributions to the effective equation of state are identified.

\subsection{Evolution equations}
If we assume a flat Friedmann-Lema\^itre-Robertson-Walker (FLRW) background space-time, the Einstein frame metric can be written as
\begin{equation}\label{flrwE}
	ds^2=-dt^2+a^2(t)d\ell^2=a^2(t)(-d\eta^2+d\ell^2)
\end{equation}
where $a(t)$ is the Einstein frame cosmic scale factor, $t$ is Einstein frame cosmic time and $\eta$ is conformal time. The corresponding Jordan frame metric can be written as

\begin{eqnarray}
	d\t s^2&=&-d\t t^2+\t a^2(t)d\ell^2=A^2(\phi)ds^2\nonumber\\
	&=&-A^2(\phi(t)) dt^2+A^2(\phi(t))a^2(t)d\ell^2\cdot
\end{eqnarray}
The observable cosmic time $\t t$ and the observable cosmic scale factor ($\t a$) are obtained from the last relations:
\begin{subequations}
\begin{eqnarray}
	d\t t & = & A(\phi(t))dt  \label{ttilde} \\
	\tilde a(\t t) & = & A(\phi(t))a(t)  \cdot    \label{atilde}
\end{eqnarray}  
\end{subequations}
Replacing metric (\ref{flrwE}) in Einstein field equations (\ref{einstein}) and using the expression of the stress-energy tensor (\ref{stressenergy}), one obtains the Friedmann and acceleration equations
\begin{subequations} \label{friedacc1}
\begin{eqnarray}
	\left(\frac{a'}{a}\right)^2 & = &\frac{4\pi}{3}\phi'^2+\frac{8\pi}{3m_p^2}a^2V(\phi)+\frac{8\pi}{3m_p^2}a^2A^4(\phi)\t \rho\qquad \label{fried1} \\
	\frac{a''}{a} &=&-\frac{4\pi}{3}\phi'^2+\frac{16\pi}{3m_p^2}a^2V(\phi)\nonumber\\
	&&\quad+\frac{4\pi}{3m_p^2}a^2A^4(\phi)(\t\rho-3\t p) \label{acc1}
\end{eqnarray}      
\end{subequations}
where a $'$ denotes a derivative with respect to conformal time $\eta$ and where $\t\rho$ and $\t p$ are the observable matter density and pressure. The Klein-Gordon equation (\ref{kleingordon}) becomes
\begin{equation}
	\phi''+2\frac{a'}{a}\phi'=-\frac{k(\phi)}{m_p^2}a^2A^4(\phi)(\t\rho-3\t p)-\frac{a^2}{m_p^2}\frac{dV}{d\phi}.\label{klein1}
\end{equation}

Finally, the conservation equation (\ref{consT}) gives
\begin{equation}        \label{erhop}
	\rho'+3\frac{a'}{a}(\rho+p)=k(\phi)\phi'(\rho-3p).
\end{equation}
Introducing the equation of state
\begin{equation} \label{state}
	\omega=\frac{p}{\rho}=\frac{\t p}{\t\rho},
\end{equation}
and replacing the Einstein frame matter density and pressure by their Jordan frame counterpart (\ref{rhop}), the conservation equation (\ref{erhop}) becomes
\begin{equation} \label{constrho}
	(A(\phi)a)^{3(1+\omega)}\t\rho=\t\rho_0
\end{equation}
where subscript $0$ refers to the actual epoch characterized by a $\tilde a_0=A(\phi_0)a_0=1$. From the last equation, we see that the Jordan frame stress-energy tensor is conserved since we find the usual expression $\tilde a^{3(1+\omega)}\t\rho=cst$.

We can introduce the observable density parameter defined by
\begin{equation}     \label{tomega}
	\t\Omega = \frac{8\pi\t G\t\rho}{3\t H^2}
\end{equation}
where $\t H$ is the observable Hubble parameter defined by 
\begin{equation} \label{htildedef}
	\t H=\frac{1}{\t a}\frac{d\t a}{d\t t}
\end{equation}                            
and $\t G$ is the varying gravitational strength which is related to the Einstein frame gravitational constant by~\cite{damour:1990fk}
\begin{equation}
	\t G=A^2(\phi)G=\frac{A^2(\phi)}{m_p^2}\cdot
\end{equation}

Replacing $\t\rho_0$ in the conservation relation (\ref{constrho}) by its value derived from (\ref{tomega}), one finds
\begin{equation}\label{trho}
	\t\rho=\frac{3m_p^2\t H_0^2\t\Omega_0}{8\pi A^2(\phi_0)(A(\phi)a)^{3(1+\omega)}}.
\end{equation}

We can now use the equation of state (\ref{state}) and the expression of $\t\rho$ (\ref{trho}) in the Friedmann, acceleration and Klein-Gordon equations (\ref{friedacc1}-\ref{klein1}). If we introduce two components in the universe: pressureless matter (representing baryonic and dark matter) characterized by $\omega_m=0$ and radiation characterized by $\omega_r=\frac{1}{3}$, we find the cosmological evolution equations
\begin{subequations}\label{cosmoequ}
\begin{eqnarray}
   \left(\frac{a'}{a}\right)^2  & = & \frac{\tilde H_0^2}{A^2(\phi_0)}\left[\frac{\t \Omega_{m0}A(\phi)}{a}\quad+ \frac{\t \Omega_{r0}}{a^2}\right] \nonumber\\
&&+\frac{4\pi}{3}\phi'^2+\frac{8\pi}{3m_p^2}a^2V(\phi)  \label{fried}\\
   	\frac{a''}{a} &=& \frac{\tilde H_0^2}{A^2(\phi_0)}\frac{\tilde\Omega_{m0}A(\phi)}{2a}\nonumber\\
&&\quad-\frac{4\pi}{3}\phi'^2+\frac{16\pi}{3m_p^2}a^2V(\phi) \label{acc}\\
 \phi''&=&-\frac{3k(\phi)}{8\pi}\frac{\t H^2_0}{A^2(\phi_0)}\frac{\t\Omega_{m0}A(\phi)}{a} \nonumber\\
&&  \quad  -2\frac{a'}{a}\phi'     -\frac{a^2}{m_p^2}\frac{dV}{d\phi}. \label{klein}
\end{eqnarray}
\end{subequations}
It is worth to comment these equations. First of all, they depend on different parameters: the present value of the densities parameters $\t\Omega_{m0}$ and $\t\Omega_{r0}$, the present value of the Hubble parameter $\t H^2_0$, the coupling function $A(\phi)$ (in this case parametrized by the value of the coupling constant $k$) and the potential $V(\phi)$ (in this case parametrized by the constants $\Lambda$ and $\alpha$). The quintessence cosmological equations are recovered in the case of a constant coupling constant ($A(\phi)=1$ and $k(\phi)=0$)~\cite{alimi:2010uq}. The above equations are somewhat different on what can be found in the literature~\cite{brax:2004fk,gannouji:2010kx} because we have decided to work with \emph{Jordan frame} density and pressure (these are directly observable) following what is done by Damour et al.~\cite{damour:1993kx,damour:1993vn,damour:1996uq} and in Babichev et al.~\cite{babichev:2010fk} while most of chameleon papers are using some hybrid conserved density ($\bar \rho$) that can be related to the traditional Einstein frame density ($\rho$ non-conserved due to the coupling to the scalar field, see~(\ref{consT})) or to the Jordan frame density pressure ($\tilde \rho$ conserved in Jordan frame)~\cite{khoury:2004uq,khoury:2004fk,brax:2004fk,gannouji:2010kx}:
\begin{equation}      \label{rhobar}
	\bar\rho=A^{-1}(\phi)\rho=A^3(\phi)\t\rho\cdot
\end{equation}  
Naively, all these definitions of matter density could be used in the source terms of the field equations. Nevertheless from an interpretative point of view, the use of the Jordan frame density $\t\rho$ is justified since Jordan frame quantities are measured as in GR with metric $\t g_{\mu\nu}$ coupling universally to matter fields. Moreover, equation (\ref{rhobar}) is not robust since it depends on symmetries of the cosmological principle and, even
in that case, is only valid for pressureless matter. Jordan Frame quantities are conserved ones, while not suffering from these pathologies, and can further safely be used in solar system physics.

\subsection{Luminosity distance} \label{sec:lum}
SNe Ia standard candles provide a measurement of the observable luminosity distance $\t d_L(\t z)$. This luminosity distance is easily obtained in the Jordan frame. Nevertheless, to illustrate the fact that observable quantities can also be derived in Einstein frame, we will derive the expression for $\t d_L$ in both frames. The present result is therefore frame independent.

In Jordan frame, the luminosity distance is expressed as in GR
\begin{equation} \label{tdl1}
	\t d_L(\t z)=(1+\t z)\int_0^{\t z}\frac{dy}{\t H(y)}
\end{equation}
where $\t z=1/\t a-1$ is the observable cosmological redshift. Finally, the distance modulus is defined as
\begin{equation}  \label{tmu1}
\t	\mu=25+5\log \left( \frac{\t d_L}{1 \ Mpc}\right)\cdot
\end{equation}
If we want to express the same observable quantity from Einstein frame, we introduce the Einstein frame distance luminosity $d_L$
\begin{equation}   \label{dl1}
	d_L(z)=(1+z)\int_0^z \frac{dy}{H(y)} 
\end{equation}
where $z=1/a-1$ and $H$ is the Einstein frame Hubble parameter $H=\frac{1}{a}\frac{da}{dt}=\frac{a'}{a^2}$. In Einstein frame, the physical units are rescaled by a factor $A(\phi)$. For this reason, the observable distance modulus is given by
\begin{equation}             \label{mu1}
	\mu=25+5\log \left(\frac{d_L}{A(\phi)\times 1 \ Mpc}\right)\cdot
\end{equation}

The equivalence between the two expressions (\ref{tmu1}) and (\ref{mu1}) can be shown. Substituting $1+z=1/a$, $H=\frac{a'}{a^2}$ and $dz=-\frac{da}{a^2}$ in (\ref{dl1}), we find
\begin{equation}
	d_L(z)=-\frac{1}{a}\int_{a_0}^a \frac{da}{a^2H}=-\frac{1}{a}\int_{\eta_0}^\eta d\eta=\frac{\eta_0-\eta}{a},
\end{equation}
and thus 
\begin{equation}\label{mu}
\mu=25+5\log\left(\frac{\eta_0-\eta}{A(\phi)a}\right)	
\end{equation} 
(if $\eta$ is expressed in $MPc$ which is possible since $c \eta$ has dimension of a length).

On the other hand, using relation (\ref{ttilde}) and the fact that $dt=ad\eta$ we can express the Hubble parameter as
\begin{displaymath} 
	\tilde H=\frac{1}{\t a}\frac{d\t a}{d\t t}=\frac{1}{\t a A(\phi)}\frac{d\t a}{dt}=\frac{\tilde a'}{\t a^2}\cdot
\end{displaymath}                                                           
Substituting this relation and using the fact that $d\tilde z=-\frac{d\t a}{\t a^2}$, the relation (\ref{tdl1}) becomes
\begin{equation}
	\t d_L(\t z)=-\frac{1}{\t a}\int_{\t a_0}^{\t a} \frac{d\t a}{\t a^2\t H}=-\frac{1}{\t a}\int_{\eta_0}^\eta d\eta=\frac{\eta_0-\eta}{\t a} \cdot
\end{equation}
Finally, the distance modulus (\ref{tmu1}) is unambiguously given by
\begin{displaymath}
\t	\mu=25+5\log\left(\frac{\eta_0-\eta}{\t a} \right) =25+5\log\left(\frac{\eta_0-\eta}{A(\phi) a} \right)
\end{displaymath}   
with $\eta$ expressed in $MPc$. Relation (\ref{atilde}) shows that the last expression is equivalent to (\ref{mu}). This shows the physical equivalence between the two frames and gives a useful formula to compute the distance modulus without any ambiguities from any frame. The distance modulus can be evaluated from the integration of the cosmological evolution equations (\ref{cosmoequ}) that gives the evolution of $a$ and $\phi$ with respect to $\eta$. 

\subsection{Supernova likelihood analysis}  \label{sec:likelihood}
In this section, we present the results of a likelihood analysis of recent SN Ia data~\cite{kowalski:2008zr}. Our goal is to identify models characterized by a set of parameters which are statistically consistent with observations, while departing from the standard $\Lambda$CDM values. Cosmological models are characterized by 6 parameters: $\t \Omega_{mo}$, $\t\Omega_{r0}$, $\t H_0$, $k$, $\Lambda$ and $\alpha$. Since the present radiation density parameter is very low, it has a negligible influence on SN Ia measurements. Therefore, we fix its value $\t\Omega_{r0}=7.97\ 10^{-5}$~\cite{kowalski:2008zr}. The value of the energy scale of the potential $\Lambda$ is optimized such that for a given value of $\alpha$, the input value of $\t \Omega_{m0}$ is retrieved. Finally, since the Hubble diagram leaves $\t H_0$ unconstrained in the plane $\alpha-\t\Omega_{m0}$~\cite{caresia:2004kx,schimd:2007vn}, we fix the value of $\t H_0=70.4\ km/s/MPc$~\cite{komatsu:2011ys}. Because of the degeneracy of the Hubble diagram with respect to $\t H_0$, we compute the marginalized likelihood over $\t H_0$~\cite{lewis:2002zr,di-pietro:2003ly}. 
\begin{figure*}[htb]
\begin{center}
\includegraphics[width=0.4\textwidth]{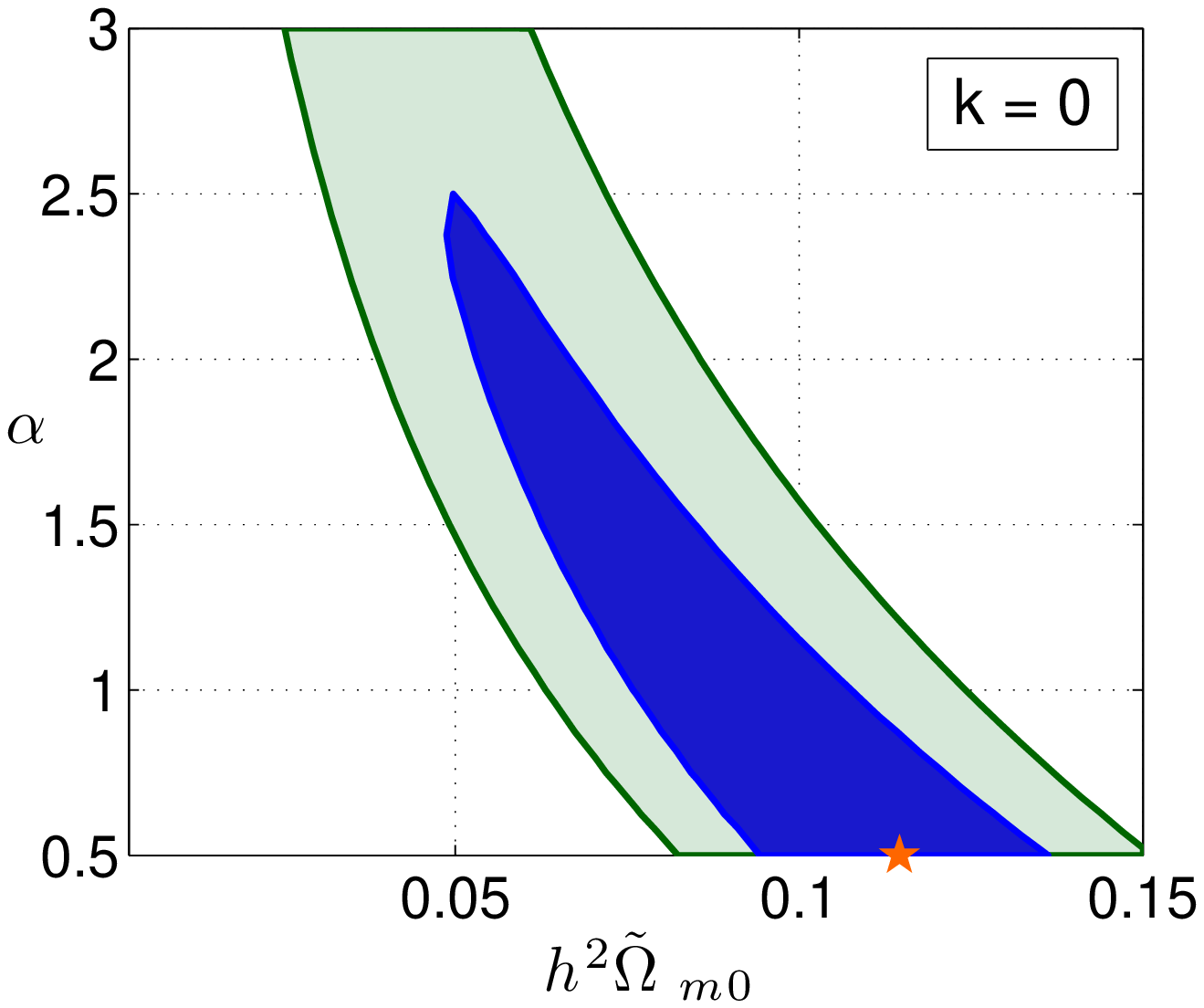}
\includegraphics[width=0.4\textwidth]{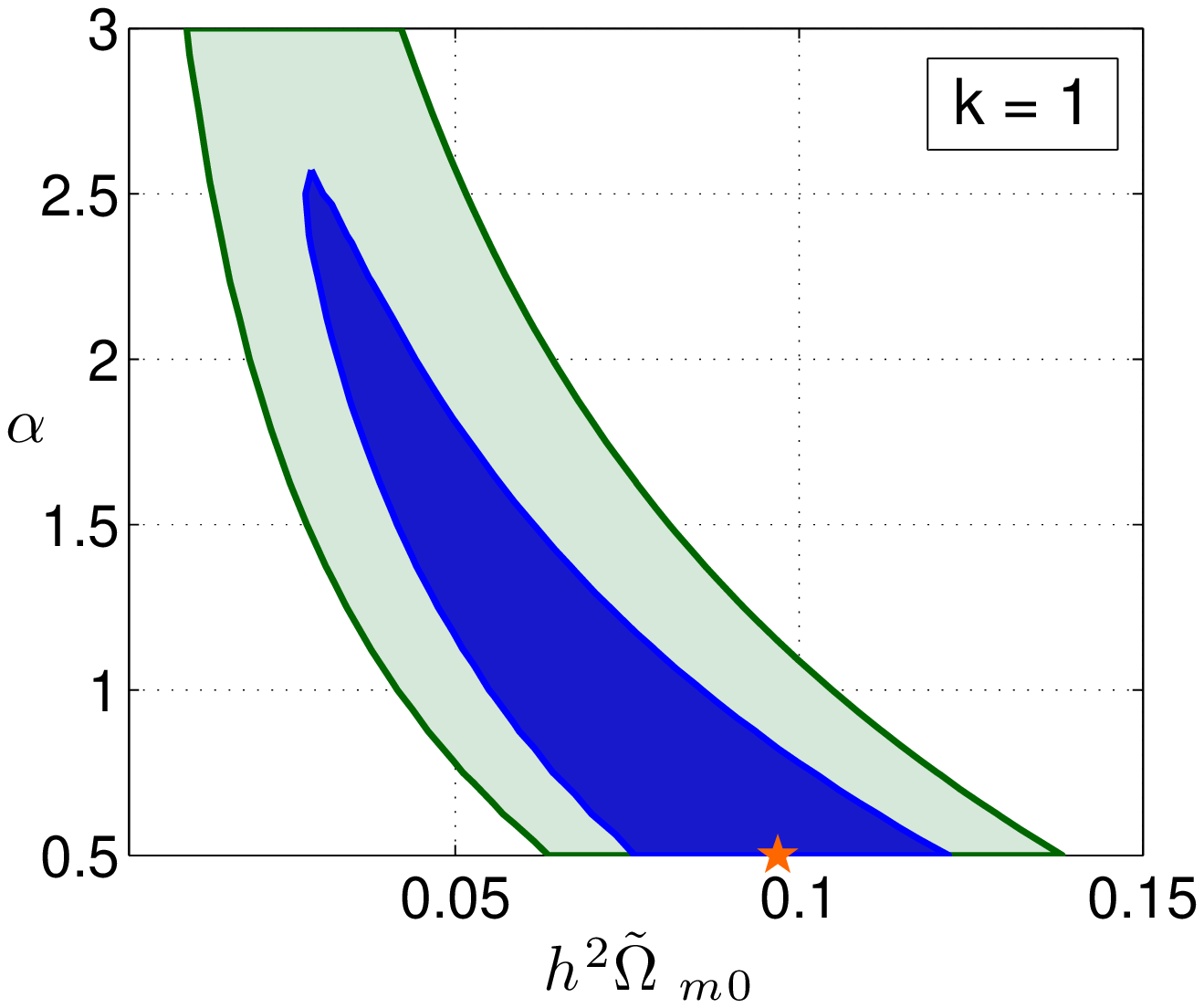}
\includegraphics[width=0.4\textwidth]{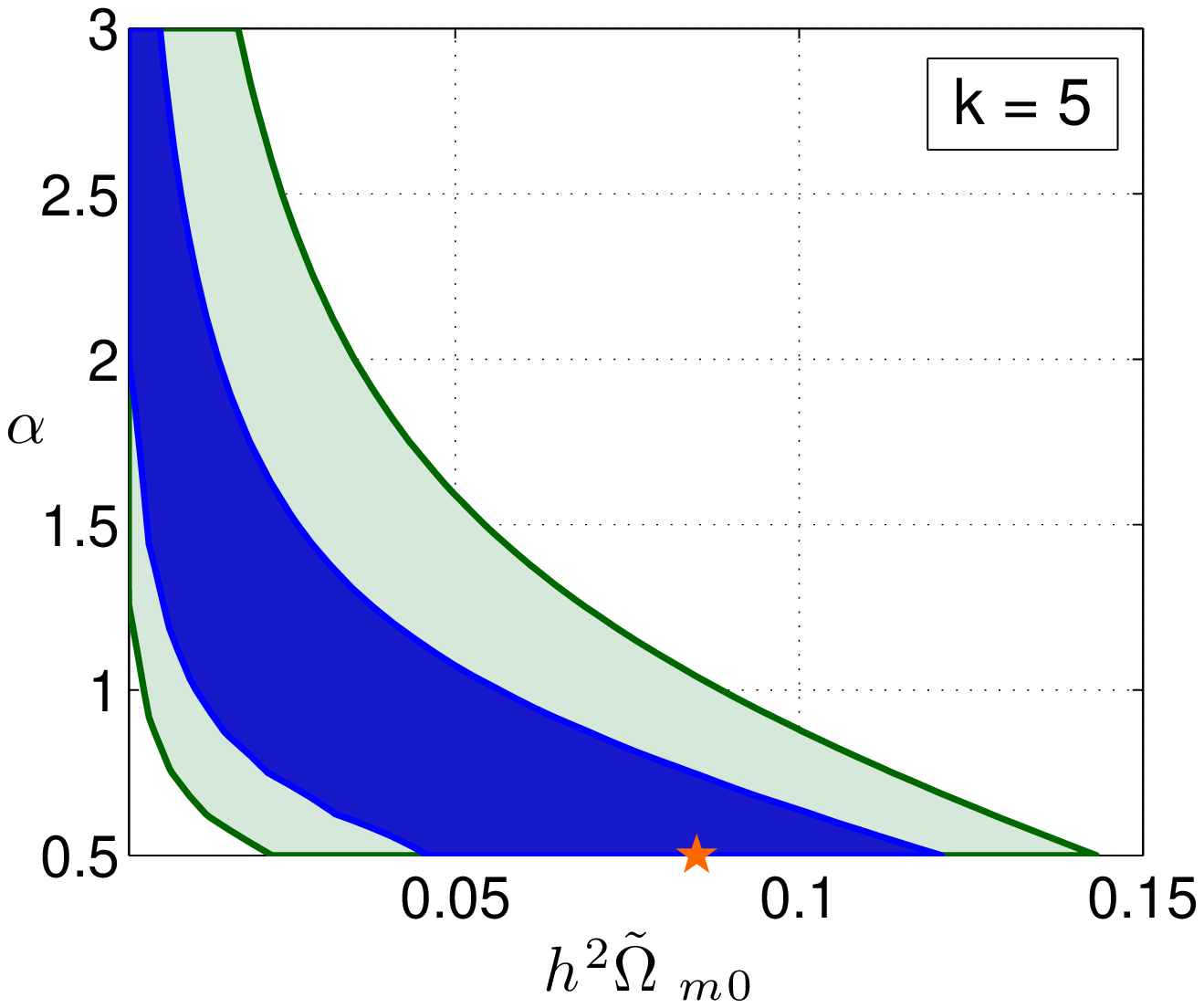}                                        
\includegraphics[width=0.4\textwidth]{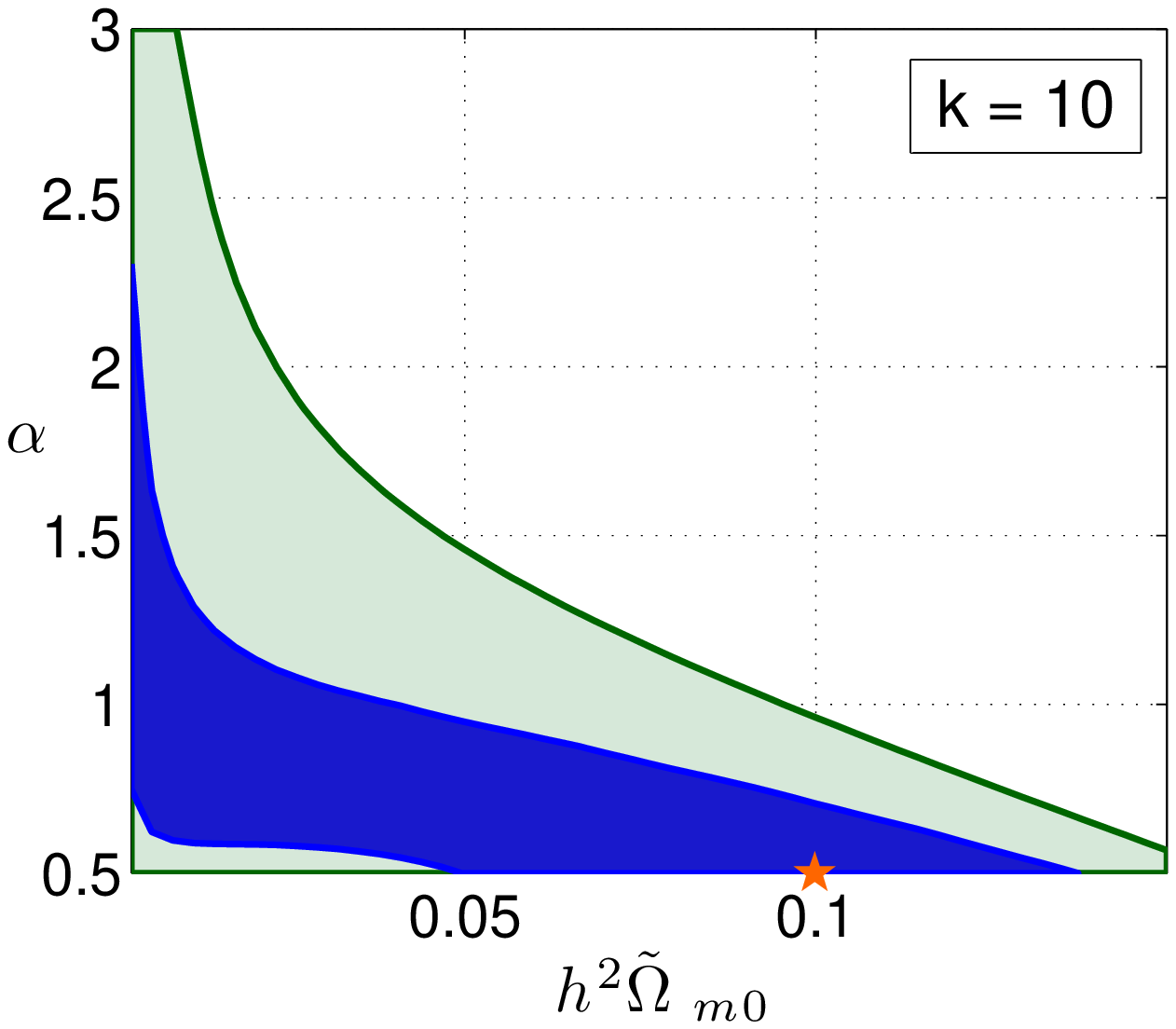}
\end{center}
\caption{Representation of the conditional 68 \% (in blue-dark) and 95\% (in green) confidence regions in the $h^2\t\Omega_{m0}-\alpha$ plane for different assumed values of coupling constant $k$ from the analysis of the UNION SN Ia Hubble diagram. The star represents the best fit.} 
\label{figTrust}
\end{figure*}

After the above considerations, each model is finally characterized by a set of three parameters ($\t\Omega_{m0}$, $\alpha$, $k$). We run a series of model decomposing the parameters domain using a uniform grid. The results of this analysis is summarized on Fig~\ref{figTrust} where we plot the conditional 68 \% and 95 \% confidence regions in the plane $\tilde\Omega_{m0}h^2-\alpha$ (with $h=\tilde H_0/100 \ km/s/Mpc$) with an assumed value of the coupling constant $k$. As can be seen, the concordance $\Lambda$CDM model (corresponding to $k=0$ and $\alpha=0$) is within this confidence region. Models characterized by $k=0$ are quintessence model. As noticed from previous quintessence data analysis~\cite{alimi:2010uq,corasaniti:2004ly}, quintessence models tend to fit the data by requiring values of $\t \Omega_{m0}h^2$ lower than in $\Lambda$CDM. This trend is reinforced when increasing the value of the coupling constant $k$. In addition, we can see that the area of the confidence region does not reduce significantly with an increase of the coupling constant $k$. This means that the fine-tuning on parameters $\alpha$ and $\t\Omega_{m0}$ does not evolve with $k$. This is mainly due to new dynamical regimes explaining cosmic expansion with a small value of the matter density parameter but a high value of the coupling constant as can be seen further.
 
\begin{figure}[htb]
\begin{center}        
	\includegraphics[width=.47\textwidth]{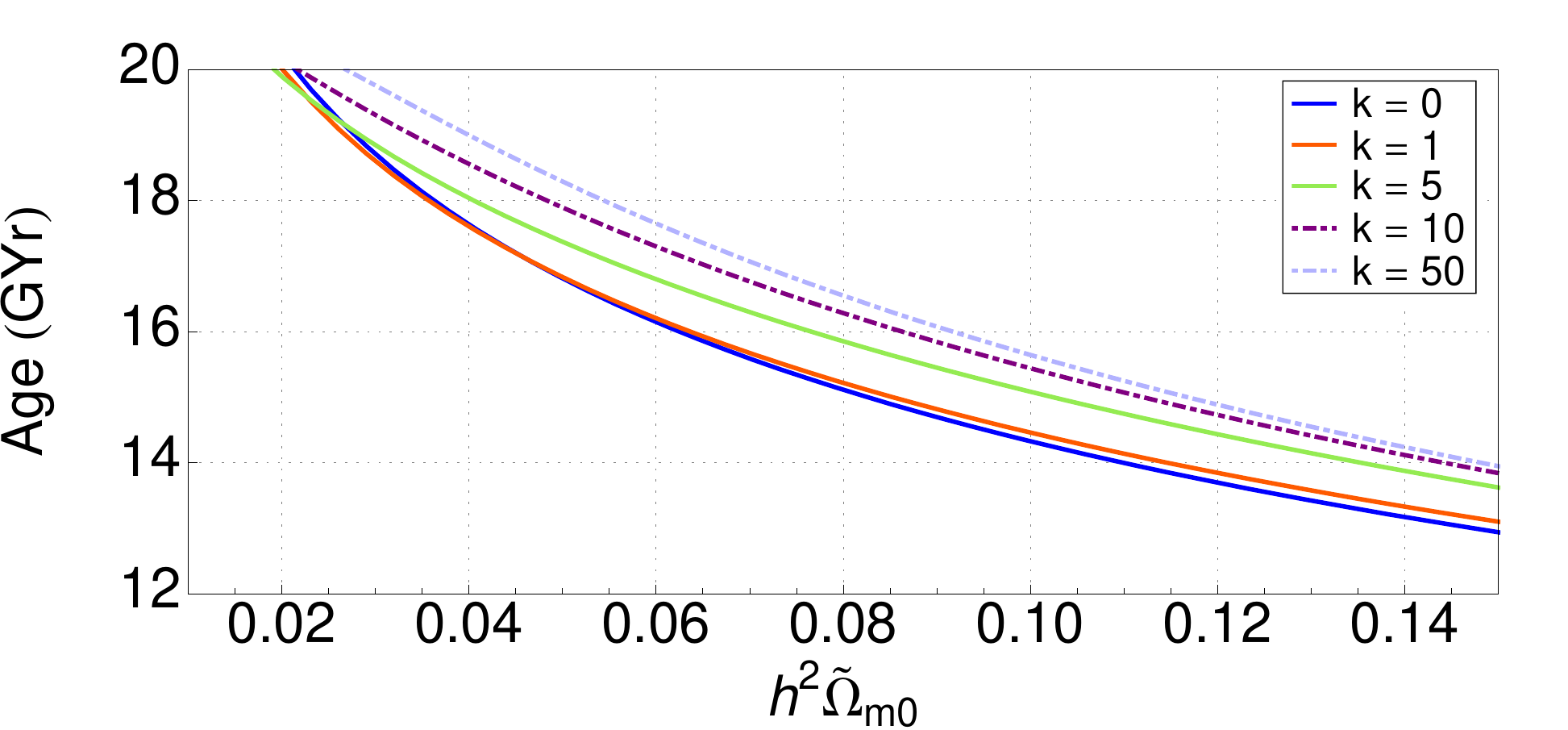}
	\includegraphics[width=.47\textwidth]{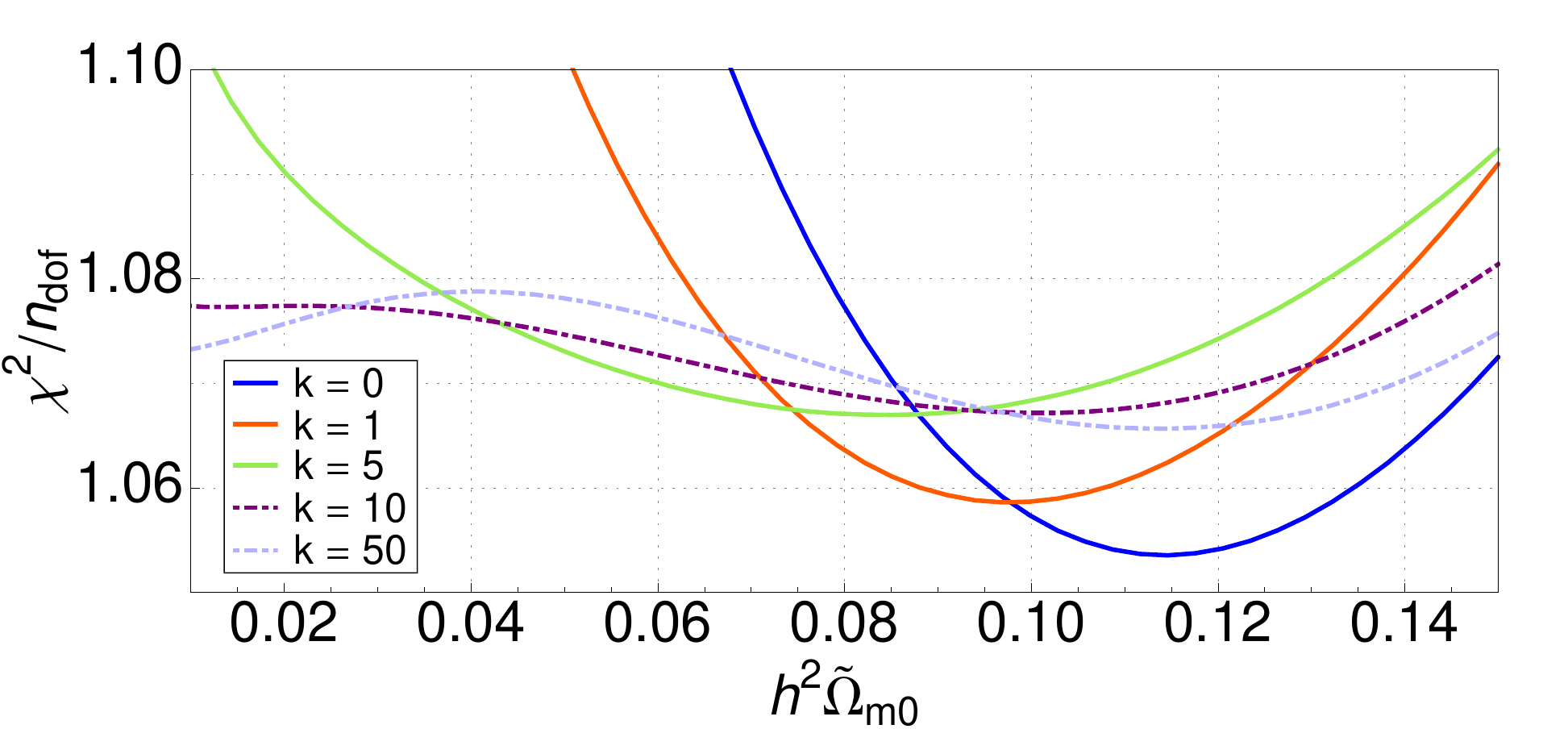}
\end{center}
\caption{Top: Representation of the age of the universe with respect to the observed matter density $\t\Omega_{m0}$ for models characterized by $\alpha=0.5$.\\
Bottom: Representation of the chi squared per degrees of freedom $\chi^2/n_{dof}$ (number of degree of freedom) with respect to the observed matter density $\t\Omega_{m0}$ for models characterized by $\alpha=0.5$.} 
\label{figTrustAge}
\end{figure}

Fig.~\ref{figTrustAge} shows the evolution of the $\chi^2$ with the observed cosmological density $\t\Omega_{m0}$ for a fixed value of $\alpha=0.5$. The $\chi^2$ gives an idea of the goodness of the fit of the model to the data. An increase of the coupling constant $k$ results in an increase of the minimum value of the $\chi^2$ which shows that models with non-vanishing coupling constant provide a slightly worst reproduction of data.   For high values of the coupling constant, the curve of the $\chi^2$ has two minima: one with a small value of $\t\Omega_{m0}$ and one with a usual value of $\t\Omega_{m0}$. These two minima characterize two very different cosmological evolutions (as it will be discussed below) and they lead to very different prediction for the age of the universe. The prediction for the age of the universe is represented on top of Fig.~\ref{figTrustAge} as a function of $\t\Omega_{m0}$. First of all, it is interesting to note that models with a high coupling constant lead to a higher estimation of the age of universe than quintessence models with the same cosmological parameters. For example for $h^2\t\Omega_{m0}=0.1$, the estimation of the age of the universe of a quintessence model is $14.2 \ 10^9$ years while for the same parameters, the model with a coupling constant $k=10$ gives an age of $15.3 \ 10^9$ years. Moreover, models with a small value of $\t\Omega_{m0}$ (allowed by the likelihood analysis) leads to an estimation of the age of the universe much higher. 

\begin{figure}[htb]
\begin{center}
\includegraphics[width=0.5\textwidth]{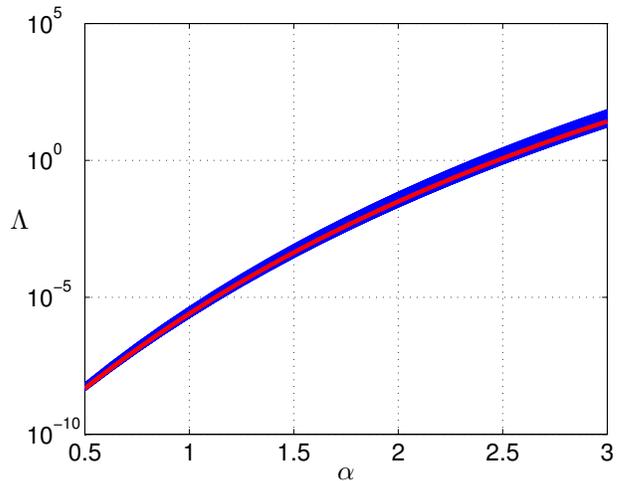}
\end{center}
\caption{In blue: data representing $\Lambda$ with respect to $\alpha$ for different value of $k$ and $\t\Omega_{m0}$. It can be seen that the influence of $k$ and $\t\Omega_{m0}$ is very weak compared to $\alpha$. In red: curve resulting from a fit on these data (see (\ref{lambdaAlpha})).} 
\label{figLambda}
\end{figure}

A supplementary interesting result is the relationship between $\Lambda$ and the other parameters. Fig. \ref{figLambda} represents the evolution of $\Lambda$ with respect to $\alpha$ for different values of $k$ and $\t\Omega_{m0}$. It can be seen that the influence of the coupling constant and of the matter density parameter is very weak in comparison with the influence of $\alpha$. We fit a curve on these data and we find the following relation
\begin{equation}  \label{lambdaAlpha}
	\log \Lambda \approx \frac{19\alpha-47}{4+\alpha}
\end{equation}                                       
which is the same relation as found in~\cite{schimd:2007vn} for quintessence model. The last relation gives an order of magnitude for the value of $\Lambda$.

\subsection{Evolution of the scalar field}\label{sec:scalarCosmo}   
It is interesting to study the cosmological evolution of the scalar field for cosmological model compatible with the SN Ie analysis from previous section. The Klein Gordon equation governing the scalar field evolution is given by (\ref{klein}) and  can be written as
\begin{eqnarray}
	\phi''+2\mathcal H\phi'&=&-k(\phi)\frac{a^2A^4(\phi)}{m_p^2}\t\rho_m-\frac{a^2}{m_p^2}\frac{dV}{d\phi}\nonumber\\
	&=&-a^2\frac{\partial V_{\rm eff}}{\partial \phi}  \label{phievol}
\end{eqnarray}
where $\mathcal H=a'/a$ and $V_{\rm eff}(\phi)$ is an effective potential whose expression is
\begin{equation} \label{veff}
	V_{\rm eff}(\phi,\t\rho_m)=\frac{1}{m_p^2}\left(V(\phi)+\frac{1}{4}A^4(\phi)\t\rho_m\right)\cdot
\end{equation}
With the hypothesis adopted in this paper (the use of a Ratra-Peebles potential and an exponential coupling function), this effective potential exhibits a minimum (see Fig.~\ref{figEffPot}). The scalar field is thus attracted toward the minimum of this effective potential. This minimum depends on the scale factor $a$ (due to the fact that the effective potential depends on $\t\rho_m$) and therefore is moving continuously with time. The shape of the effective potential for two different densities is represented in Fig.~\ref{figEffPot}. The density is decreasing with the scale factor (see equation (\ref{constrho})) which implies this effective potential was very narrow in the early time and it becomes wider and wider with the time.          

It is worth noticing that the expression (\ref{veff}) of the effective potential is somewhat different of the expression given in Brax et al~\cite{brax:2004fk}. The difference comes only from the fact that we use the Jordan frame density $\t\rho$ while Brax et al used a hybrid definition of the density $\bar \rho$ given by (\ref{rhobar}).

Working with Ratra-Peebles potential $V(\phi)=\Lambda^{4+\alpha}/m_p^\alpha\phi^\alpha$ and with an exponential coupling function $A(\phi)=e^{k\phi}$ gives an analytical expression for the minimum of the effective potential determined by the following conditions
\begin{equation}
	0=-\frac{\alpha\Lambda^{4+\alpha}}{m_p^\alpha\phi^{\alpha+1}}+k\t\rho_m e^{4k\phi}\cdot
\end{equation}                                                             
Solving this equation gives the value of the scalar field that minimizes the effective potential $\phi_{min}$ as a function of the cosmological density $\t\rho_m$, of the coupling constant $k$ and of the parameters of the potential ($\alpha$ and $\Lambda$):
\begin{equation}                           \label{phimin}
	\phi_{min}(\t\rho_m)=\frac{\alpha+1}{4k}\ LW\left(\frac{4kb}{\alpha+1}\right)
\end{equation}                                 
with $b=\left(\frac{\alpha\Lambda^{4+\alpha}}{m_p^\alpha k\t\rho_m}\right)^{\frac{1}{\alpha+1}}$  and $LW(x)$ the W-Lambert function~\cite{chapeau-blondeau:2002fk}.

\begin{figure}[htb]
\begin{center}
\includegraphics[width=0.47\textwidth]{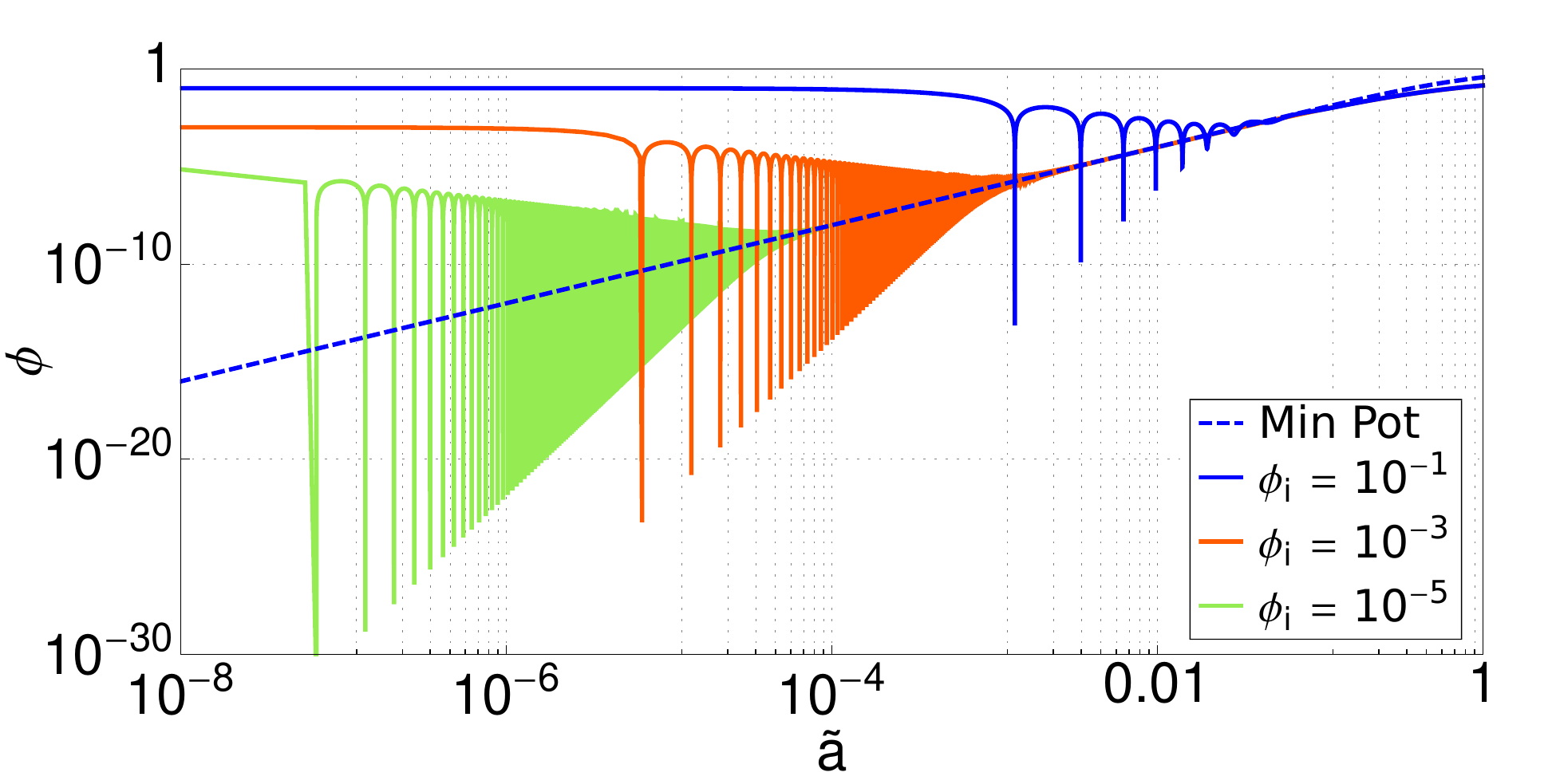}
\includegraphics[width=0.47\textwidth]{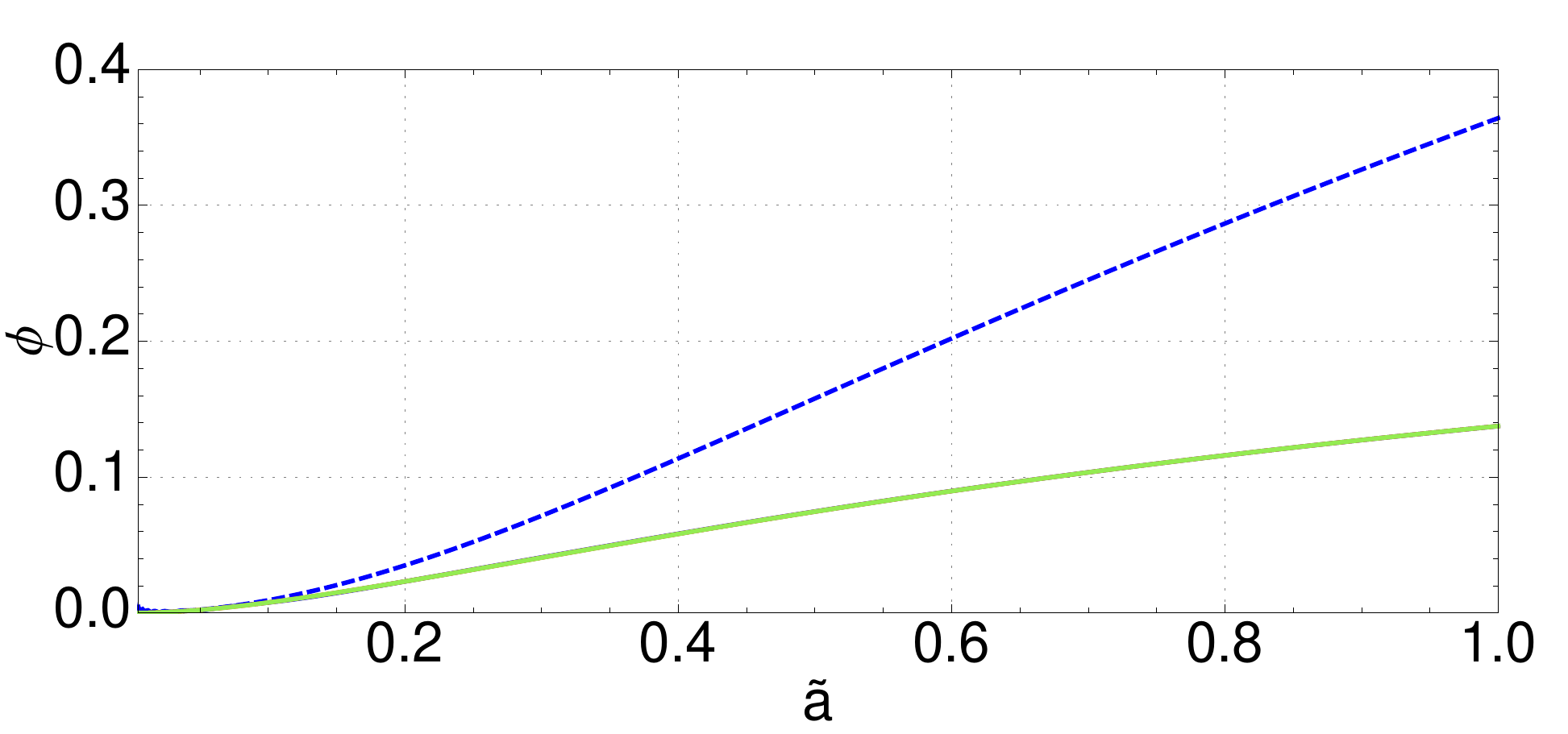}
\end{center}
\caption{Evolution of the scalar field as a function of the Jordan frame scale factor $\tilde a$ (the observable one) for different initial conditions $\phi_i$ and for the model characterized by ($k=1$, $\t\Omega_{m0}=0.197$ and $\alpha=0.5$). The minimum of the effective potential (see relation (\ref{phimin})) is represented in blue. On top: curves on a log-log scale, we can see that the scalar field reaches the minimum of the effective potential for all the initial conditions. On bottom: linear x-axis, we see that the scalar field does not follow the minimum of the effective potential from $\tilde a \sim 0.1$ and the evolution of the scalar field is independent of the initial condition $\phi_i$ (all curves are superimposed).} 
\label{figPhiMin}
\end{figure}

\begin{figure}[htb]
\begin{center}
\includegraphics[width=0.47\textwidth]{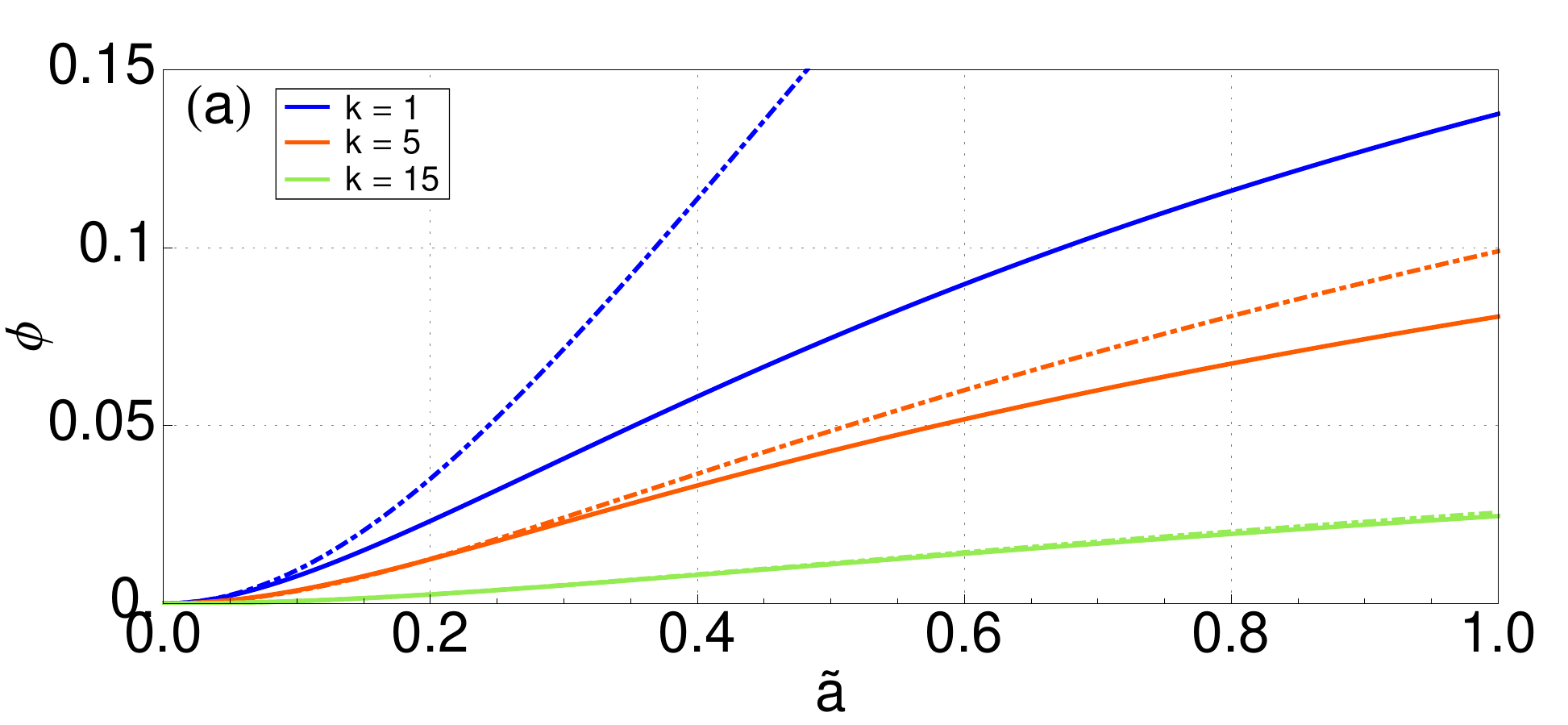}
\includegraphics[width=0.47\textwidth]{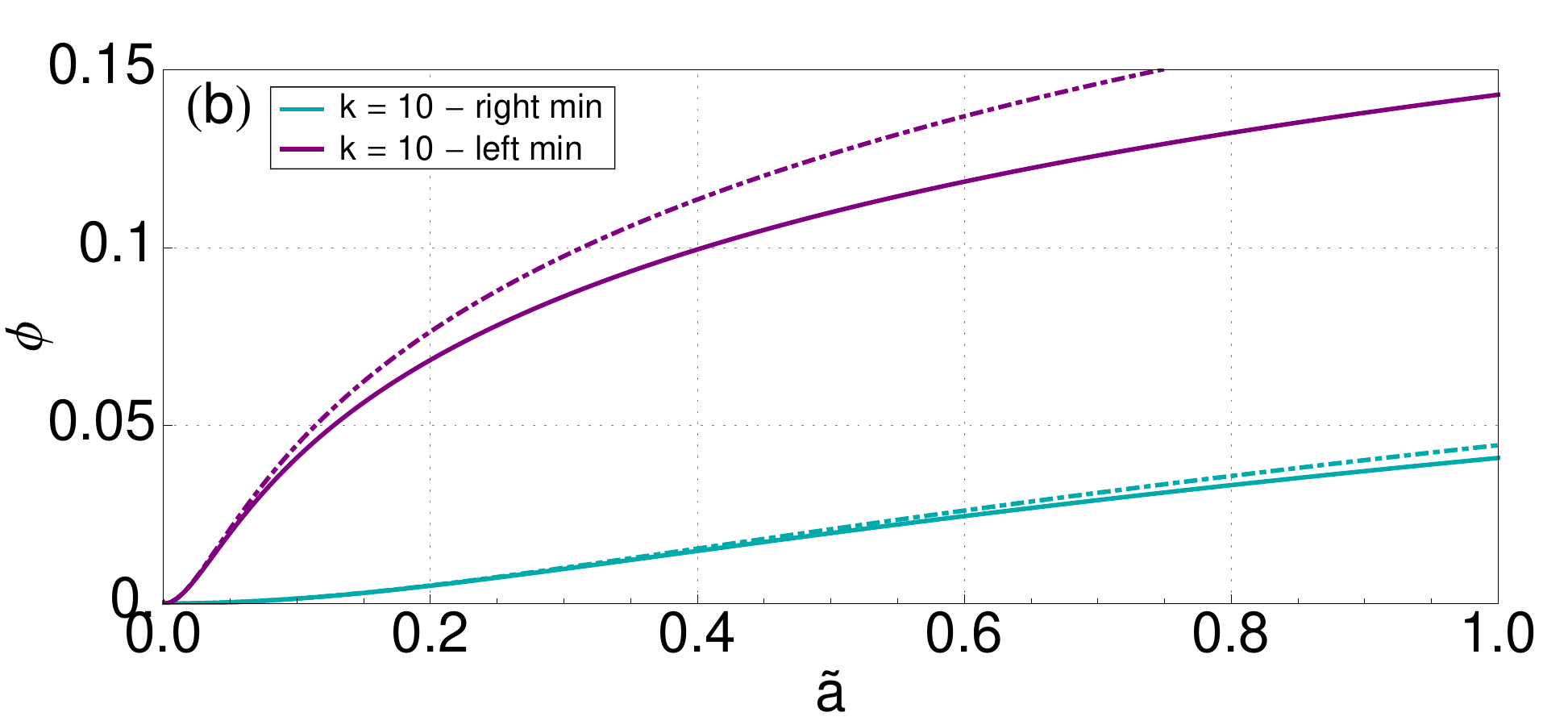}
\hspace*{1mm}\includegraphics[width=0.462\textwidth]{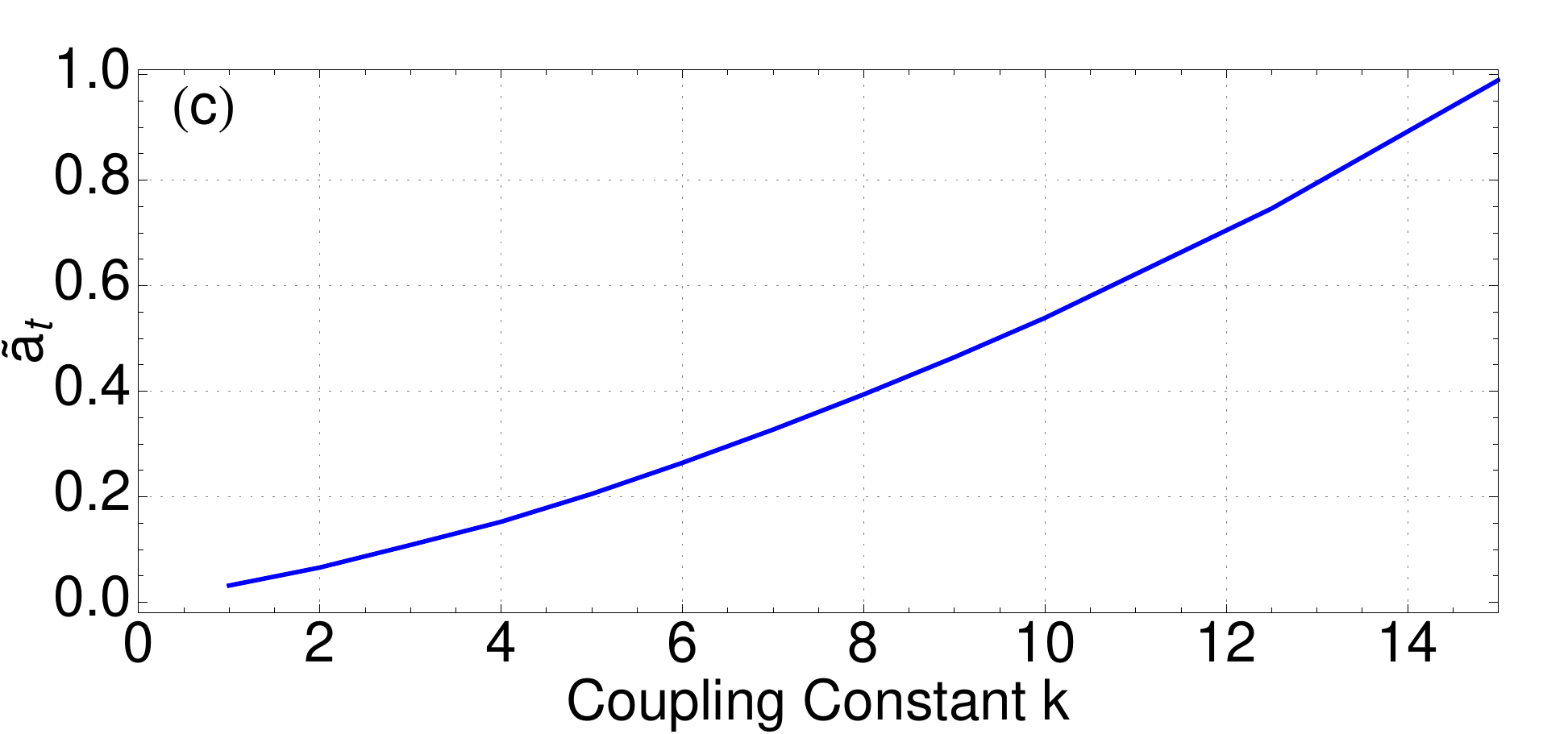}
\end{center}
\caption{(a) : Evolution of the scalar field (continuous lines) and of the minimum of the effective potential $\phi_{min}$ (\ref{phimin}) (dashed lines) for different values of the coupling constant $k$. For small values of the coupling constant, the scalar field leaves $\phi_{min}$ quite early. The transition scale factor $\t a_{t}$ where the field leaves the potential increases with the coupling constant $k$. For $k=15$, the scalar field is still in the minimum of the effective potential at the present epoch. \\                                                                                                                
(b) : Evolution of the scalar field (continuous lines) and of the minimum of the effective potential $\phi_{min}$ (\ref{phimin}) (dashed lines) for two models with $k=10$ located in the local minima of the $\chi^2$ (see Fig.~\ref{figTrustAge}). The evolution of the model located in the second minimum of the $\chi^2$ located in the low $\t\Omega_{m0}$ minimum is very fast.\\
(c): Evolution of the transition scale factor $\t a_t$ with the coupling constant $k$.} 
\label{figPhiEvol}
\end{figure}

For models fitting the SNe Ia (see previous section), the typical behavior of the scalar field if the following:

\begin{itemize}
	\item at the beginning, the attractor mechanism is very efficient and the scalar field is attracted by the minimum of the effective potential and oscillates around it. This is due to the fact that the effective potential is very narrow in the early time. Fig. \ref{figPhiMin} shows the behavior of the scalar field for different initial conditions and the minimum of the effective potential $\phi_{min}$ (\ref{phimin}). We can clearly see that the attractor mechanism is very strong for small scale factor. This implies that the cosmological evolution is very weakly sensitive to the scalar field initial conditions (see Fig.~\ref{figPhiMin}). This is a reminiscence of the tracking properties of the Ratra-Peebles potential~\cite{steinhardt:1999fk}.
	\item during some time, the scalar field follows closely the minimum of the effective potential.
	\item depending on the model considered, the attractor mechanism may be not strong enough and the scalar field may not be able to follow the minimum of the potential (see Fig.~\ref{figPhiMin} and Fig.~\ref{figPhiEvol}). In particular for low coupling constants, the effective potential becomes too wide to attract sufficiently the scalar field. 
\end{itemize}

\begin{figure}[htb]
\begin{center}
\includegraphics[width=0.48\textwidth]{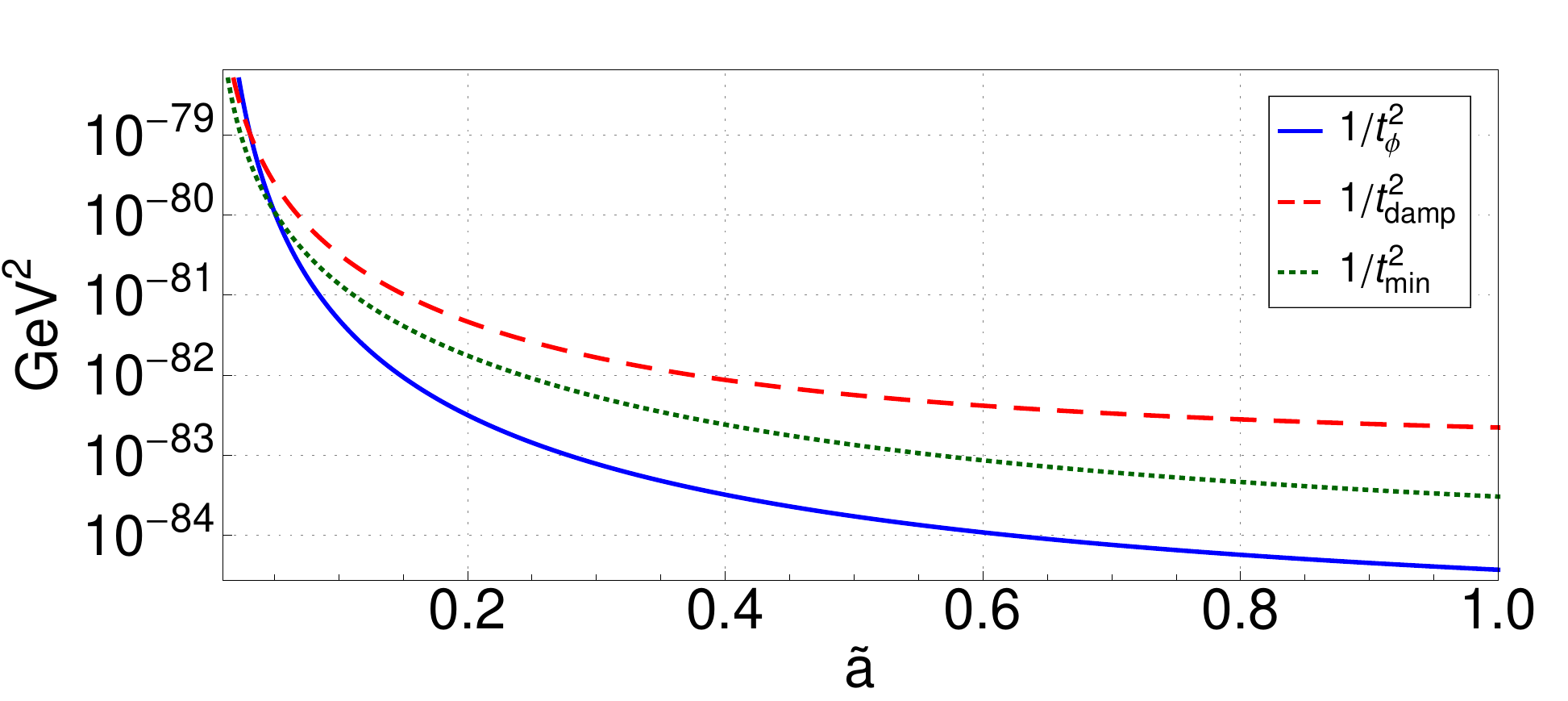}
\end{center}             
\caption{Evolution of the three time scales characterizing the field response time (\ref{tphi}), the Hubble damping (\ref{tdamp}) and the evolution of the effective potential (\ref{tmin}) as a function of the Jordan frame cosmic scale factor $\t a$ for the same model as in Fig.~\ref{figPhiMin}. We can see the intersection between $t_\phi$ and $t_{damp}$ occurred at the transition scale factor $\t a_t$ that corresponds to the scale factor where the field leaves the minimum of the potential (see Fig.~\ref{figPhiMin}).}
\label{figPhiTime}
\end{figure}

The necessary condition for the scalar field to follow the minimum of the effective potential can be derived by considering the time scales involved in the scalar field evolution. There are three different time scales:
\begin{itemize}
	\item the damping time-scale due to the $2\mathcal H\phi'$ in equation (\ref{phievol}) term given by
	\begin{equation} \label{tdamp}
		\frac{1}{t_{damp}}=\frac{2\mathcal H}{a}=2H
	\end{equation}                                 
	where $t_{damp}$ is an Einstein frame cosmic time-scale characterizing the Hubble damping.
	
	\item the time-scale characterizing the scalar field oscillations around the minimum of the potential. This time scale is related to the effective mass of the scalar field given by
	\begin{equation}
		m^2=\frac{\partial^2 V_{\rm eff}}{\partial \phi^2}\cdot
	\end{equation}
	The related time-scale is given by
	\begin{equation}\label{tphi}
		\frac{1}{t_\phi^2}=\frac{m^2}{(2\pi)^2}
	\end{equation}
	which characterizes the response time of the scalar field.
	
	\item the last time-scale involved in the scalar field evolution is related to the evolution of the effective potential. This potential is continuously moving in time due to the presence of the $\t\rho_m$ term in (\ref{veff}). We characterize the time-scale due to this evolution by the time scale related to the variation of the minimum of the potential $\phi_{min}$
	\begin{equation}\label{tmin}
		\frac{1}{t_{min}}=\frac{\dot\phi_{min}}{\phi_{min}}=\frac{1}{\phi_{min}}\frac{\partial \phi_{min}}{\partial \t\rho_m}\frac{\partial\t\rho_m}{\partial\t a}\dot{\t a}
	\end{equation}
	where a dot denotes the derivative with respect to Einstein frame cosmic time ($dt$).
	Using the expression of $\phi_{min}$ (\ref{phimin}), of $\t\rho_m$ (\ref{trho}) and the definition of $\t a$ (\ref{atilde}), we finally get
	\begin{equation}
		\frac{1}{t_{min}}=\frac{3\left(H+\alpha\dot\phi \right)}{1+\alpha+4k\phi_{min}}\cdot
	\end{equation}
\end{itemize}
The scalar field $\phi$ follows the minimum of the effective potential if the damping time is higher than the scalar field response time ($t_{damp}>t_\phi$) and if the evolution of the effective potential is slower than the field response time ($t_{min}>t_{\phi}$). In other words, the conditions characterizing the fact that the scalar field follows the minimum of the potential are given by
\begin{subequations}
\begin{eqnarray}
	\frac{m^2}{4\pi^2} & >  & 4H^2  \\
	\frac{m^2}{4\pi^2} & >  & \frac{9\left(H+\alpha\dot\phi \right)^2}{(1+\alpha+4k\phi_{min})^2} 
\end{eqnarray}     
\end{subequations}
The three time-scales are represented on Fig.~\ref{figPhiTime} (for a model fitting SN Ia data with $k=1$). We can see that the intersection of $t_{\phi}$ with the other curves occurred at $\t a\sim 0.1$ which is exactly the scale factor where the scalar field stops to follow the minimum of the effective potential (see Fig. \ref{figPhiMin}). We note $\t a_t$ the Jordan frame scale factor from which the scalar field stops to follow the minimum of $V_{\rm eff}$. Fig.~\ref{figPhiEvol} (a) represents the evolution of the scalar field and of the minimum of the effective potential for different values of the coupling constant $k$. The transition scale factor $\t a_t$ increases with $k$ and for $k\sim 15$, the scalar field stays in the minimum of the potential during all the evolution. In Fig.~\ref{figPhiEvol} (b), the behavior of the scalar field for the second minimum of the $\chi^2$ curves for $k=10$ is also represented. The evolution of this model (characterized by a small value of the density parameter $\t\Omega_{m0}$) is very different from the other ones. The scalar field increases very fast.

Fig.~\ref{figPhiEvol} (c) represents also the evolution of the transition scale factor with the coupling constant $k$ ($\t a_t$ being computed by the comparison of the Hubble damping time scale and the field response time). The transition scale factor increases with the coupling constant which is logical since the effective potential is  tighter for high coupling constant, it is therefore more difficult to leave this potential. 

\subsection{Evolution of the cosmic expansion} \label{sec:cosmic}
We study here what is the origin of the observed cosmic acceleration. In this context, cosmic expansion is described in the Jordan frame by the scale factor $\t a=A(\phi)a$ that is measured with matter. In order to identify the origin of the acceleration, we need to compute the value of $\frac{1}{\t a}\frac{d^2\t a}{d\t t^2}$. With the definition of $\t a$ (\ref{atilde}) and with the evolution equations (\ref{cosmoequ}), we can express
\begin{equation} \label{tacc}
  \t q= \frac{1}{\t H^2 \t a}\frac{d^2\t a}{d\t t^2}=\t q_{M} + \t q_{Q} +\t q_{\rm NMC}
\end{equation}
where all the terms are given by
\begin{subequations} \label{taccd}
\begin{eqnarray}
	\t q_{M}&=&-\frac{\t \Omega_m}{2}-\t \Omega_r \\
	\t q_{Q}	&=&\frac{8\pi}{3m^2_pA^2(\phi)\t H^2}\left(V(\phi)-m_p^2\frac{\phi'^2}{a^2}\right)  \\
	 \t q_{\rm NMC}&=&\frac{k\phi''}{A^2(\phi)a^2\t H^2}\cdot
\end{eqnarray}
\end{subequations}
The first term gives the usual term present in $GR$ involving matter (here the subscript $M$ refers to presure-less baryonic and dark matter and also to radiation). The second term $\t q_{Q}$ is a term due to the presence of the scalar field in the dynamics. This term is also found in quintessence models (where the coupling function is equal to $A(\phi)=1$ and the coupling constant $k$ vanishes)~\cite{steinhardt:1999fk,di-pietro:2003ly,schimd:2007vn}. Finally, the last term $\t q_{\rm NMC}$ is due to the non-minimal coupling ($A(\phi)\neq cst$) and is closely related to the evolution of the scalar field. 

\begin{figure}[htb]
\begin{center}
\includegraphics[width=0.48\textwidth,clip=true,trim=17 27 10 0]{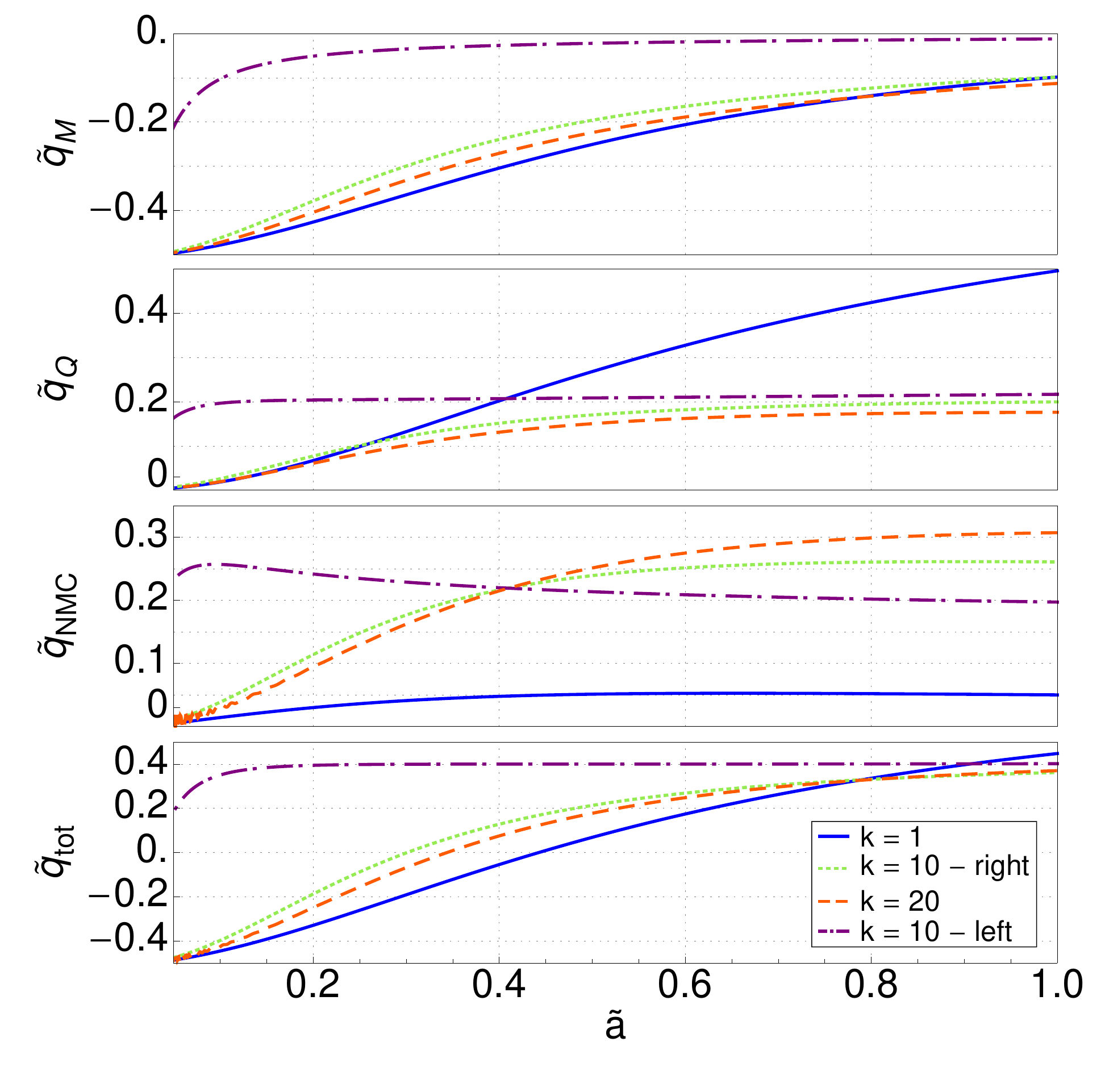}                                
\end{center}
\caption{Representation of the different terms of the acceleration factor for different coupling constant $k$. The different terms $\tilde q_M$ for the matter contribution, $\tilde q_Q$ for the quintessence contribution and $\tilde q_{\rm NMC}$ for the non minimal coupling contribution are given by equations (\ref{taccd}). The three first curves $k=1,10,20$ represents model in the global minimum of the $\chi^2$ while the last curve $k=10 {\rm - left}$ represents a model in the second minimum (left) of the $\chi^2$ (see Fig.~\ref{figTrustAge}).  For model in the global minimum of the $\chi^2$: the GR contribution is of the same order of magnitude for different values of $k$; the quintessence potential contribution decreases when the coupling constant increases to leave the room for the contribution coming from the non-minimal coupling. The model in the low $\t\Omega_{m0}$ minimum of $\chi^2$ (left) characterized by a lower value of the $\t\Omega_{m0}$ has a much faster evolution. }
\label{figqtilde}
\end{figure}

The different contributions to the cosmic acceleration are represented on Fig.~\ref{figqtilde} for three different values of the coupling constant $k$ ($k=1$, $k=10$ and $k=20$) for best-fit model. The matter contribution is more or less the same for the different $k$. What is interesting to notice is that for small value of $k$, the cosmic acceleration is explained by the quintessence term $\t q_Q$ while the non-minimal coupling does not play any role. An increase of the coupling constant produces a decrease of the quintessence contribution in favor of the non-minimal coupling contribution $\t q_{\rm NMC}$. In conclusion, if $k>>1$, the cosmic acceleration is explained by the non-minimal coupling.

In order to compare the difference in the dynamic of the two minima in the $\chi^2$ curves (see Fig.~\ref{figTrustAge} (bottom) and discussion related), Fig.~\ref{figqtilde} represents also the evolution of the acceleration factors for the two different minimum for $k=10$. The evolution of the model in the left-minimum (with small matter density parameter) is very fast and reaches an asymptotic behavior with the quintessence contribution being of the same order of magnitude than the contribution from the non minimal coupling. The matter contribution tends to a very small value which is logical since the density parameter of this model is very small.
%
%
%

\subsection{Dark-Energy effective equation of state}
Following what is usually done in quintessence studies~\cite{steinhardt:1999fk,di-pietro:2003ly,schimd:2007vn,alimi:2010uq,alam:2003fk} and in the context of the AWE hypothesis~\cite{alimi:2008zr}, we can modeled the presence of the non-minimally coupled scalar field by an effective fluid of dark energy in GR (parametrized by a density parameter $\t \Omega_{DE}$ and by the equation of state parameter $\t\omega_{DE}$). In order to identify clearly the quintessence contribution from the contribution coming from the non minimal coupling, we introduce two effective fluids in GR related to quintessence term (characterized by $\t \Omega_Q$ and $\t \omega_Q$) and to the non minimal coupling (characterized by $\t \Omega_{\rm NMC}$ and $\t \omega_{\rm NMC}$).

The observed Hubble constant is defined by (\ref{htildedef}). Using relations (\ref{atilde}) and (\ref{ttilde}), we can relate this Jordan frame Hubble constant to the Einstein frame one
\begin{equation}     \label{htilde}
	\t H=\frac{1}{\t a}\frac{d\t a}{d\t t}=\frac{1}{Aa}\frac{d\left(Aa\right)}{Adt}=\frac{1}{A}\left(H+\alpha (\phi)\frac{d\phi}{dt}\right)
\end{equation}

\begin{figure}[htb]
\begin{center}
\hspace*{2mm}\includegraphics[width=0.46\textwidth]{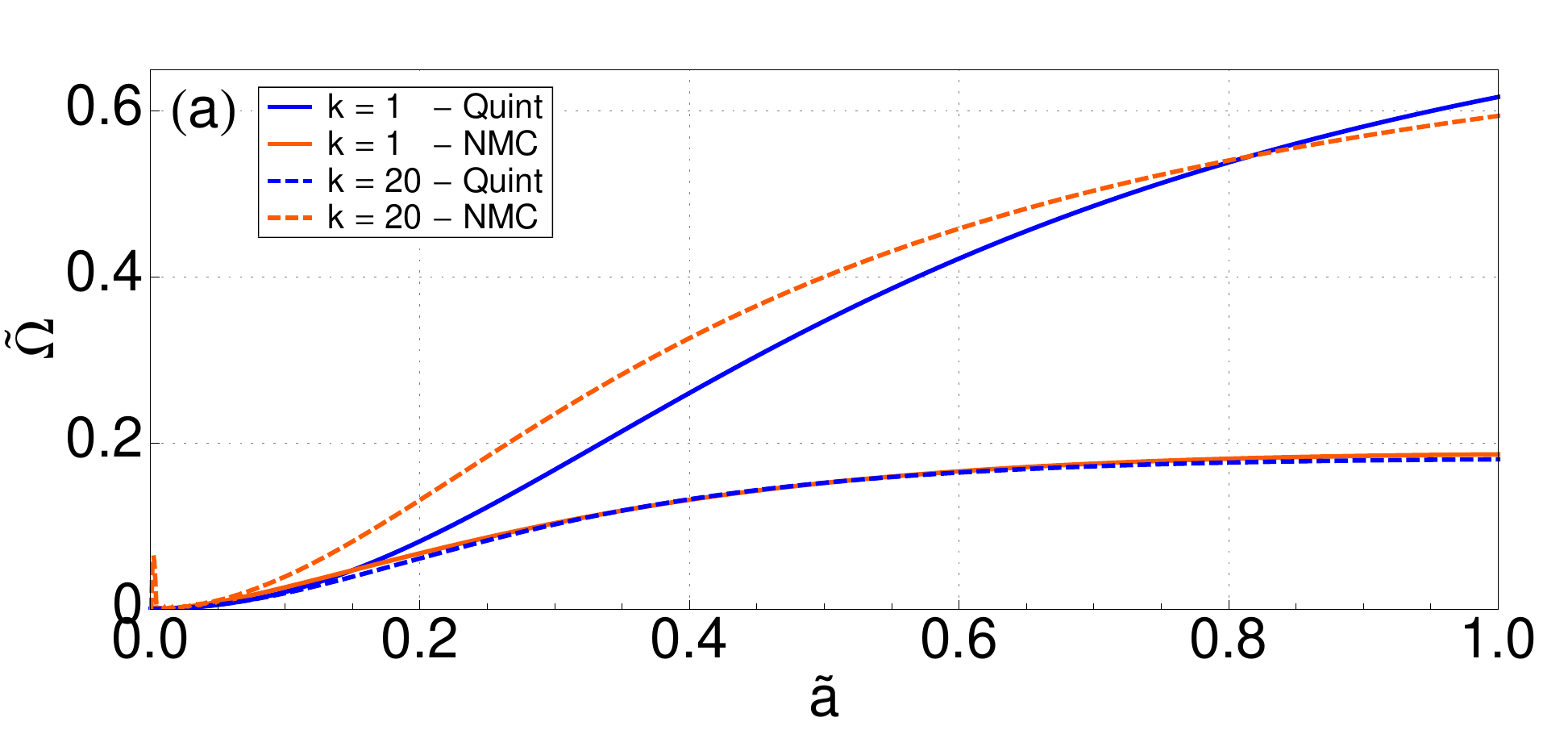} 
\hspace*{2mm}\includegraphics[width=0.46\textwidth]{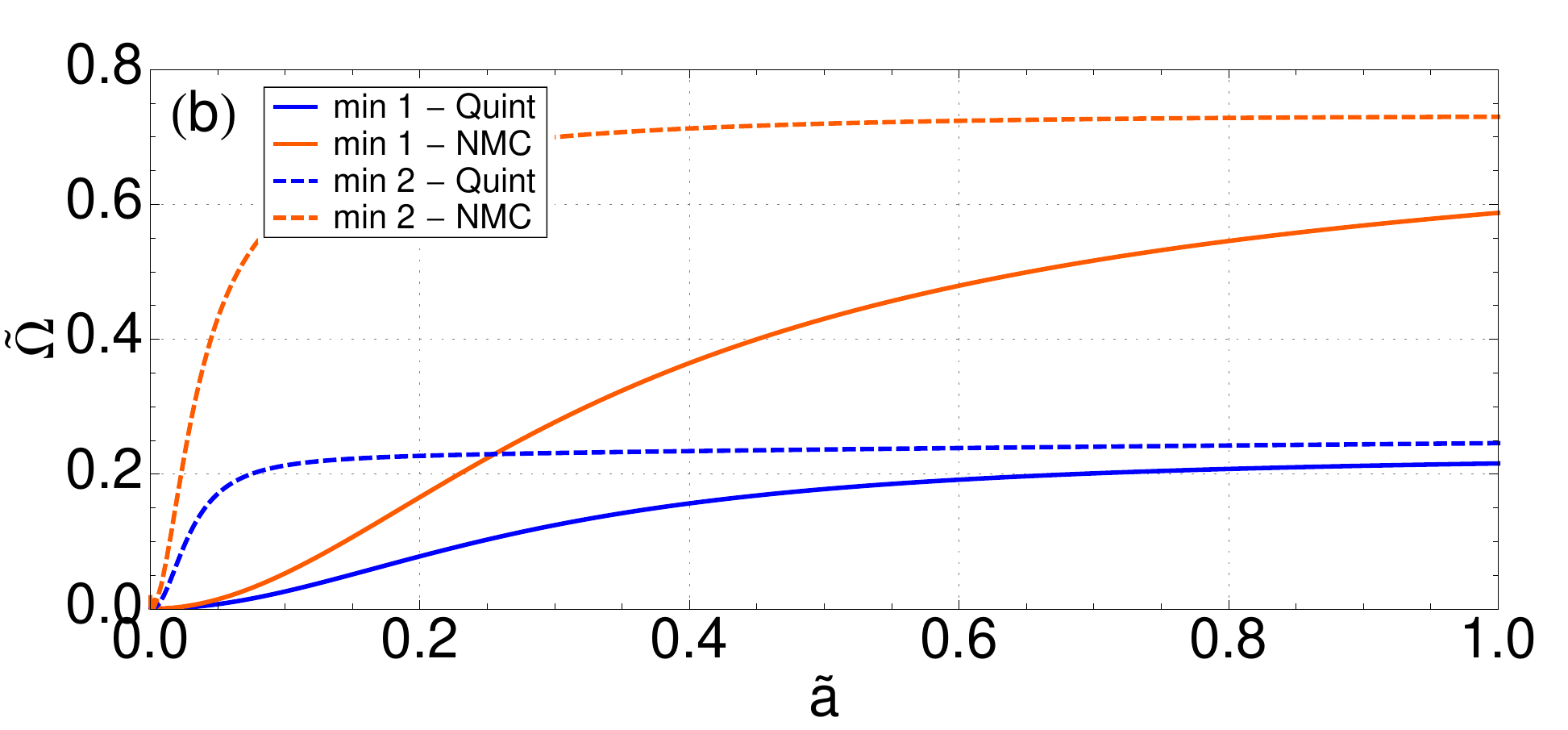}     
\includegraphics[width=0.47\textwidth]{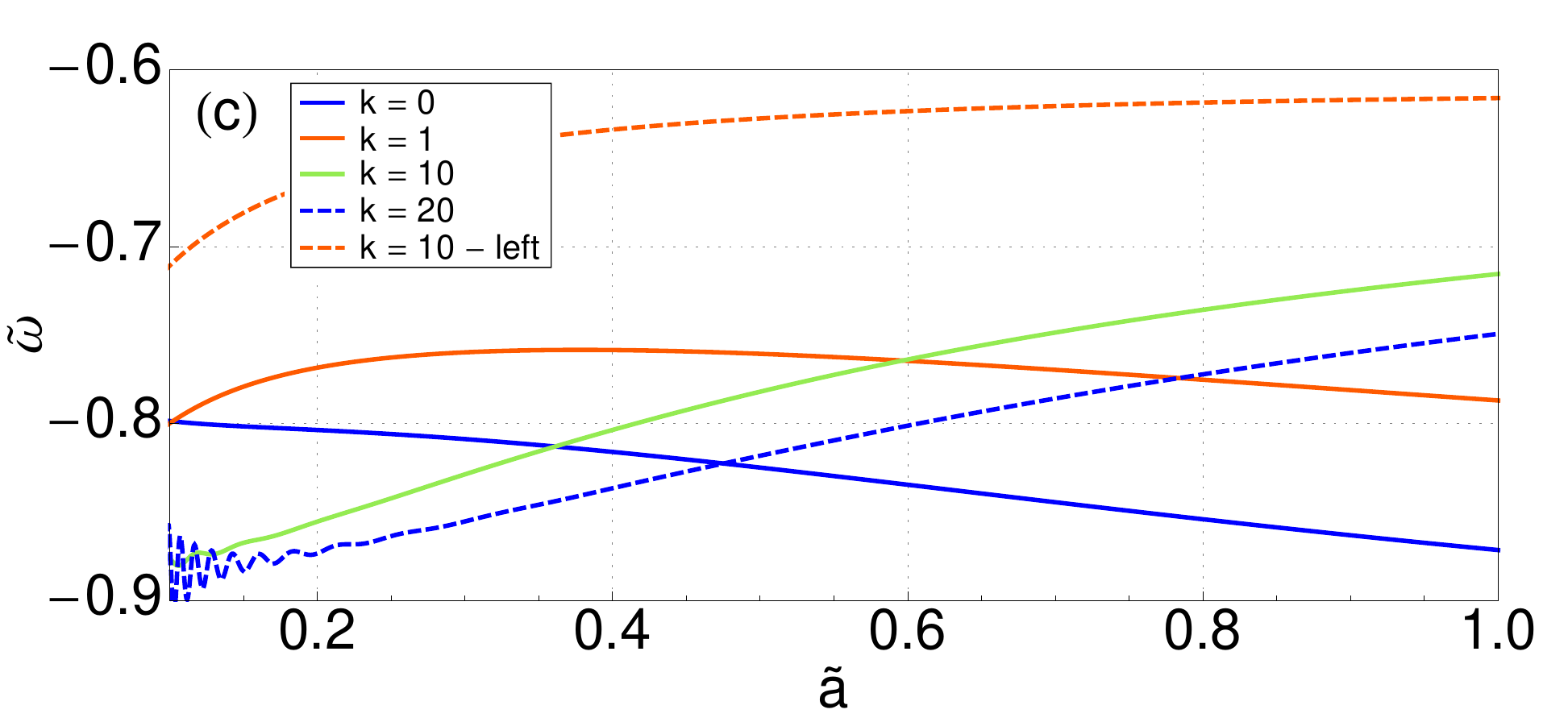} 
\end{center}
\caption{(a-b) : Evolution of the density parameter $\t \Omega$ corresponding to the quintessence contribution ($\tilde \Omega_Q$ given by (\ref{omegaq}) and represented in blue (dark)) and to the non-minimal coupling contribution ($\tilde \Omega_{\rm NMC}$ given by (\ref{omeganmc}) and represented in red (light)) different coupling constant: Fig. (a) : $k=1$ and $k=20$ ; Fig. (b): two models characterized with $k=10$ and in the different local minima of the $\chi^2$\\ 
(c) : evolution of the equation of state parameter $\t \omega_{DE}$ as a function of the cosmic scale factor for best-fit model with different coupling constants and for a model in the second minimum of the $\chi^2$ with $k=10$.} 
\label{figomega}
\end{figure}

 Inserting the evolution equations (\ref{cosmoequ}) in (\ref{htilde}) and by matching it with the usual Friedmann equations in the Jordan frame
\begin{equation}
	1=\t\Omega_{m}+\t \Omega_r+\t \Omega_Q +\t \Omega_{\rm NMC},  
\end{equation}
one identifies the expression of the observable density parameters 
\begin{subequations}
\begin{eqnarray}
	\t\Omega_Q &=  & \frac{8\pi}{3m_p^2A^2(\phi)\t H^2}\left(\frac{m_p^2\dot \phi^2}{2}+V(\phi) \right) \\
	\t \Omega_{\rm NMC}&=&\frac{k}{\t  H^2A^2(\phi)}\left(2\frac{\dot a\dot \phi}{a}+k\dot \phi^2\right)\cdot
\end{eqnarray}     
\end{subequations}
The total parameter density is given by
\begin{equation}
	\t\Omega_{DE}=\t\Omega_Q+\t\Omega_{\rm NMC}\cdot
\end{equation}   
Fig.~\ref{figomega} (a-b) represents the evolution of the different density parameters $\t\Omega$ defined above for different coupling constants $k$. The quintessence density parameter decreases when $k$ increases while the non-minimal coupling density parameter increases with $k$. This is consistent with the behavior of the acceleration factor $\t q$ illustrated in Fig.~\ref{figqtilde}. For high value of the coupling constant, the cosmic acceleration is mainly explained by the non-minimal coupling. The asymptotic behavior of the left-minimum model (low value of $\t\Omega_{m0}$) is also confirmed by Fig.~\ref{figomega} (b).

Matching the acceleration equation (\ref{tacc}) with the usual GR acceleration
\begin{displaymath}
	\tilde q=   -\frac{\t \Omega_m}{2}-\t \Omega_r-\frac{1}{2}\t\Omega_Q(1+3\t\omega_Q)-\frac{1}{2}\t\Omega_{\rm NMC}(1+3\t\omega_{\rm NMC})
\end{displaymath} 
gives the equations of state
\begin{subequations}
\begin{eqnarray}
	\t\omega_Q & =  & \frac{m_p^2\dot \phi^2/2 -V(\phi)}{m_p^2\dot \phi^2/2 +V(\phi)}  \label{omegaq}\\  
	\t\omega_{\rm NMC}&=&\frac{1}{3}\frac{2\frac{\dot a\dot \phi}{a}+2\frac{\partial V_{\rm eff}}{\partial \phi}-k\dot \phi^2 }{2\frac{\dot a\dot \phi}{a}+k\dot \phi^2}\label{omeganmc}
\end{eqnarray}                                                                                                                                                  
\end{subequations}
and the total equation of state is given by
\begin{equation}
	\t\omega_{DE}=\frac{\t\omega_Q\t\Omega_Q+\t\omega_{\rm NMC}\t\Omega_{\rm NMC}}{\t\Omega_{DE}} \cdot
\end{equation}     
Fig.~\ref{figomega} (c) represents the evolution of the total equation of state parameter $\t\omega_{DE}$ for different coupling constants $k$. All models represented on this figure are statistically equivalent to $\Lambda$CDM for the SNe Ia measurements. An increase of the coupling constant produce a significative change in the shape of the curves. In particular, the value of $\left.\frac{d\t\omega}{d\t a}\right|_0$ can become positive (this derivative is sometimes noted $-\t\omega_1$ in a Chevallier-Polarski-Linder parametrization of the equation of state parameter $\t\omega(\t a)=\t\omega_0 +\t\omega_1(1-\t a) $~\cite{chevallier:2001uq,linder:2003kx}). In quintessence scenario with Ratra-Peebles potential, the derivative of $\t\omega$ is always negative ($\t \omega_1>0$)~\cite{weller:2002vn}. Therefore, a measurement of the value of the derivative of the equation of state parameter should in principle discriminate strongly non minimally coupled DE and uncoupled quintessence. Such a measurement will be performed by the EUCLID mission recently selected by ESA.

%
%
%
%
\section{Solar system constraints} \label{ref:solarsystem} 
In this part, we will study the constraints on the parameters characterizing self interaction of the scalar field and on the coupling between matter and the scalar field that can be obtained from solar system experiments. Traditional tensor-scalar (traditional in the sense where no potential is considered or chameleon mechanism is not playing any role) is severely constrained by solar system and by binary pulsar experiments. For example, deviations from General Relativity can be expressed in the Parametrized-Post-Newtonian (PPN) framework \cite{will:1993fk,will:2006cq}. The PPN parameters related to a traditional tensor-scalar theory of gravity with an exponential coupling constant $A(\phi)=e^{k\phi}$ are given by (see e.g. \cite{damour:1992ys})\footnote{\label{footphi}Note that our scalar field $\phi$ is related to the scalar field $\varphi$ used by Damour and Esposito-Far\`ese by the relation $\phi=\frac{\varphi}{\sqrt{4\pi}}$. Therefore, our coupling constant $k$ is related to their constant $\alpha$ by $k=\sqrt{4\pi}\alpha$ and the $\t\gamma$ parameter given by $\t\gamma=\frac{1-\alpha^2}{1+\alpha^2}$ becomes $\t\gamma=\frac{4\pi-k^2}{4\pi+k^2}$}
\begin{equation}   \label{ppn1}
	\t\gamma-1 =-2\frac{k^2}{4\pi+k^2},\qquad \t\beta-1=0\cdot
\end{equation}
The present constraint on the $\t\gamma$ parameter is obtained from the Shapiro measurement of the Cassini probe~\cite{bertotti:2003uq}:
\begin{equation}  \label{cassini}
	\t\gamma-1=\left(2.1\pm 2.3\right)\times 10^{-5} \cdot
\end{equation}
This constraint on $\t\gamma$ shows that  tensor-scalar theory of gravity can be viable only for very small linear coupling parametrized by $k$. The main point of the chameleon fields presented in Khoury and Weltman~\cite{khoury:2004uq,khoury:2004fk} is that this constraint can be evaded thanks to the chameleon mechanism. If the potential of the scalar field is chosen so that its effective mass becomes large in presence of matter, the coupling constant $k$ is replaced by an effective coupling constant $k_{\rm eff}$ that can be strongly reduced with respect to $k$. The goal of this part of the paper is to study a simplified model of Sun and to derive constraints so that the chameleon mechanism is strong enough to pass the PPN test of gravity. Furthermore,  we will compare the admissible parameters region with the confidence region obtained by a cosmological analysis in the previous section.
\\

Let us summarize the innovative points of this publication for what concerns solar system physics. In the next section, we study properties related to spherical static solutions of field equations (\ref{einstein}-\ref{kleingordon}-\ref{consT}). The field equations deriving from a Schwarzschild-like metric is derived. These equations are similar that the ones found in~\cite{babichev:2010fk} but the problem of the boundary conditions is fully discussed in the light of the cosmological analysis from the previous section. In Sec.~\ref{sec:cham}, we review the chameleon mechanism presented in~\cite{khoury:2004fk,khoury:2004uq}. Sec.~\ref{sec:num} is devoted to the analysis of the analytical field profile derived in the literature \cite{khoury:2004fk,waterhouse:2006kx,tamaki:2008ve,tsujikawa:2009qf}. In particular, the hypothesis done to obtain these analytical results are discussed and a comparison between the analytical results and the full numerical resolution of the field equations is performed in order to identify the validity regime of the analytical solution. 

In Sec.~\ref{sec:ppn}, we derive the post-newtonian parameters related to the solar system metric. Different PPN analysis of the chameleon mechanism can be found in the literature~\cite{khoury:2004fk,faulkner:2007kx,capozziello:2008vn}. Unfortunately, the present results are derived using different gauge which can be misleading. Therefore, we present a complete derivation of the Post-Newtonian Parameters following the procedure described in Damour en Esposito-Far\`ese~\cite{damour:1992ys} and we compare it to current results. Finally, we combine the cosmological analysis and the Post-Newtonian analysis to derive constraints on the parameters of the theory. In particular, we show that the considered model cannot explain cosmic expansion and satisfy solar-system constraints at the same time.

\subsection{Static spherical configuration}
The metric characterizing a static spherically symmetric space-time can be written in Schwarzschild coordinates (in Einstein frame)
\begin{equation}
	ds^2=-e^{\nu(r)}dt^2+e^{\lambda(r)}dr^2+r^2d\Omega^2\cdot
\end{equation}
Replacing this expression of the metric in the field equations and using the definition of the stress-energy tensor (\ref{stressenergy}), we find the following equations
\begin{subequations}\label{fieldSun}
\begin{eqnarray}
	\lambda'&=& \frac{8\pi}{m_p^2}rA^4(\phi)e^\lambda\t\rho+4\pi r\Psi^2+\frac{1-e^\lambda}{r}\nonumber\\
	&&\qquad+\frac{8\pi}{m_p^2}re^\lambda V(\phi) \label{lambda_} \\
	\nu'&=&  \frac{8\pi}{m_p^2}rA^4(\phi)e^\lambda\t p+4\pi r\Psi^2- \frac{1-e^\lambda}{r}\nonumber\\ 
	&&\qquad-  \frac{8\pi}{m_p^2}re^\lambda V(\phi) \\
	\phi'&=&\Psi \\
	\Psi'&=& -\left(\frac{2}{r}+\frac{1}{2}(\nu'-\lambda')\right)\Psi+\frac{A^4(\phi)e^\lambda}{m_p^2} k(\phi) (\t\rho-3\t p)\nonumber\\
	&&\qquad +\frac{e^\lambda}{m_p^2}\frac{dV}{d\phi}\label{psi_}\\ 
	\t p'&=&-(\t\rho+\t p)\left(k(\phi)\Psi+\frac{\nu'}{2}\right)\label{p_}\cdot
\end{eqnarray}      
\end{subequations}                                                         
where a prime denotes here the derivative with respect to the radial coordinate $r$. These equations extend the ones derived by Damour and Esposito-Far\`ese~\cite{damour:1993vn} (where no potential was considered) and are equivalent to the ones in Babichev and Langlois~\cite{babichev:2010bh}. This set of 5 equations with 6 unknowns needs to be completed by an equation of state. In this communication, we consider a simple model of star/planet assuming the energy density is constant inside the body ($\t\rho_b$). Outside the body, we consider a cosmological background of baryonic gas and dark matter whose constant density $\t\rho_\infty$ is related to the cosmological density parameter $\t\Omega_{m0}$ by the relation (\ref{trho}). 

To understand the chameleon effect, it is useful to consider the equations (\ref{psi_}) in the non-relativistic limit (ignoring the backreaction of the scalar field and considering $e^\lambda\sim 1$) and neglecting the pressure with respect to the energy density:
\begin{equation}
	\frac{d^2\phi}{dr^2}+\frac{2}{r}\frac{d\phi}{dr}=\frac{d V_{\rm eff}}{d \phi}
\end{equation}
where the effective potential is exactly the one appearing in the cosmological evolution~(\ref{veff}).
With a runaway potential, this effective potential has a minimum given by $\phi_{min}(\t\rho)$. In the case of a Ratra-Peebles potential, this minimum is given by the expression (\ref{phimin}).    As already stated in Section \ref{sec:scalarCosmo}, the expression of the effective potential is different from what is usually used in the chameleon literature~\cite{khoury:2004uq,khoury:2004fk,tamaki:2008ve,tsujikawa:2009qf,de-felice:2010uq,brax:2004fk} because we use the always conserved Jordan frame density $\t\rho$ that are observable (we follow what is done by Damour et al.~\cite{damour:1993kx,damour:1993vn,damour:1996uq} and by Babichev and Langlois~\cite{babichev:2010fk}) while most chameleon papers are using an hybrid density for pressureless matter that is conserved in cosmological context ($\bar \rho$) related to the Jordan frame or to the Einstein frame density by a conformal factor as indicated by the relation (\ref{rhobar}). This difference does not change qualitatively the results obtained.

\begin{figure}[htb]
\begin{center}
\includegraphics[width=0.48\textwidth]{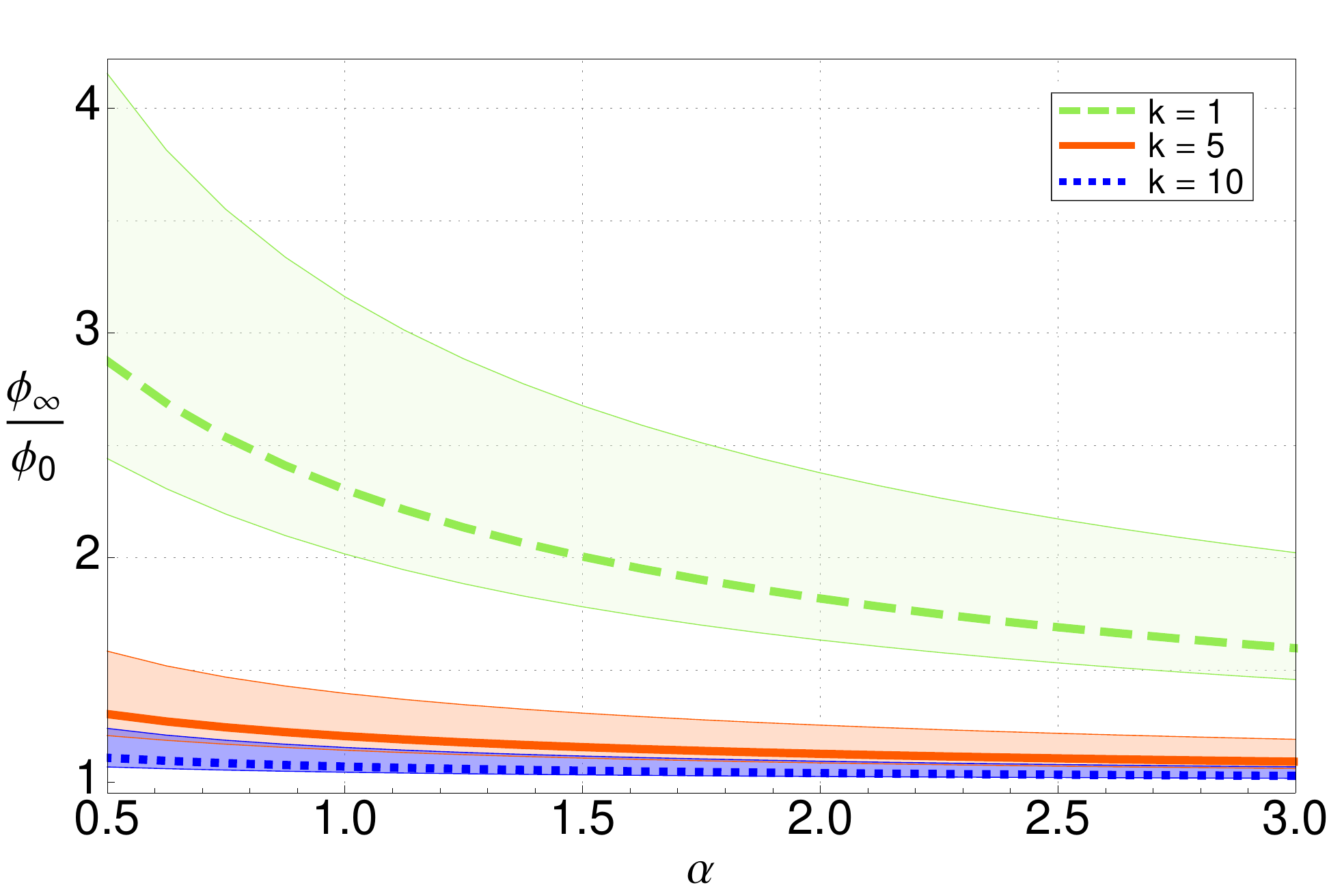}
\end{center}
\caption{Ratio $\phi_{min}(\t \rho_\infty)/\phi_0$ between the value of $\phi$ minimizing the cosmological effective potential $V_{\rm eff}$ and $\phi_0$ the value of the field given by the cosmological evolution. The different lines correspond to different values of the coupling constant $k$ and the filled areas represent different values of $\t\Omega_{m0}$. } 
\label{figPhiinf_phi0}
\end{figure}

Five boundary conditions are needed to solve the problem given by equations (\ref{fieldSun}) and the equation of state. Regularity conditions at the origin of coordinates $r=0$ impose that
\begin{subequations}\label{boundary}
\begin{eqnarray}
	\lambda(r=0)&=&0 \\
	\Psi(r=0)&=&0\cdot 
\end{eqnarray}
The matter pressure needs to vanish at the boundary of the body considered $r=R_b$, giving the supplementary condition:
\begin{equation}
	p(r=R_b)=0\cdot
\end{equation}
An additional condition on $\nu$ is still needed, which corresponds to fixing the time coordinate. This choice can be arbitrary and does not change the result of the integration since the equations depend only on $\nu'$.

The last condition concerns the scalar $\phi$. Following what is done in Khoury and Weltman~\cite{khoury:2004uq,khoury:2004fk}, in Babichev and Langlois~\cite{babichev:2010bh}, in Tamaki and Tsujikawa~ \cite{tamaki:2008ve} and in Tsujikawa et al~\cite{tsujikawa:2009qf}, we suppose that the scalar field reaches the minimum of the potential at very large distance of the body:
\begin{equation} \label{phi_inf}
	\phi(r=\infty)=\phi_{min}\cdot
\end{equation}      
\end{subequations}
The validity of the assumption is questionable. Indeed, we have seen in section \ref{sec:scalarCosmo} that the cosmological evolution does not always produce a situation where the scalar field stays in the minimum of the effective potential. In particular, for low values of the coupling constant ($k$ below $\sim 15$), we have shown that this is not the case and the cosmological value of the scalar field at the present epoch $\phi_0$ is different from the value minimizing the effective potential ($\phi_{min}$) as illustrated in Fig.~\ref{figPhiEvol}. In Fig.~\ref{figPhiinf_phi0} we represent the ratio $\phi_\infty/\phi_0$ (with $\phi_\infty$ being the value of the scalar field minimizing the effective potential). It can be seen that for low value of the coupling constant, this ratio is different from 1 and the boundary conditions (\ref{phi_inf}) can not be fullfilled. To be completely consistent, one should solve the cosmological evolution of the structures in order to have an idea of the field boundary conditions at the limit of the solar system. This work is out of the scope of this paper. 

The five equations  (\ref{fieldSun}) with the considered equation of state and with the five boundary conditions (\ref{boundary}) described above form a Boundary Value Problem (BVP). This BVP has been solved numerically by using multigrid methods.

\subsection{The chameleon mechanism}  \label{sec:cham}
\begin{figure}[htb]
\begin{center}
\includegraphics[width=0.5\textwidth,clip=true,trim=10 0 30 0]{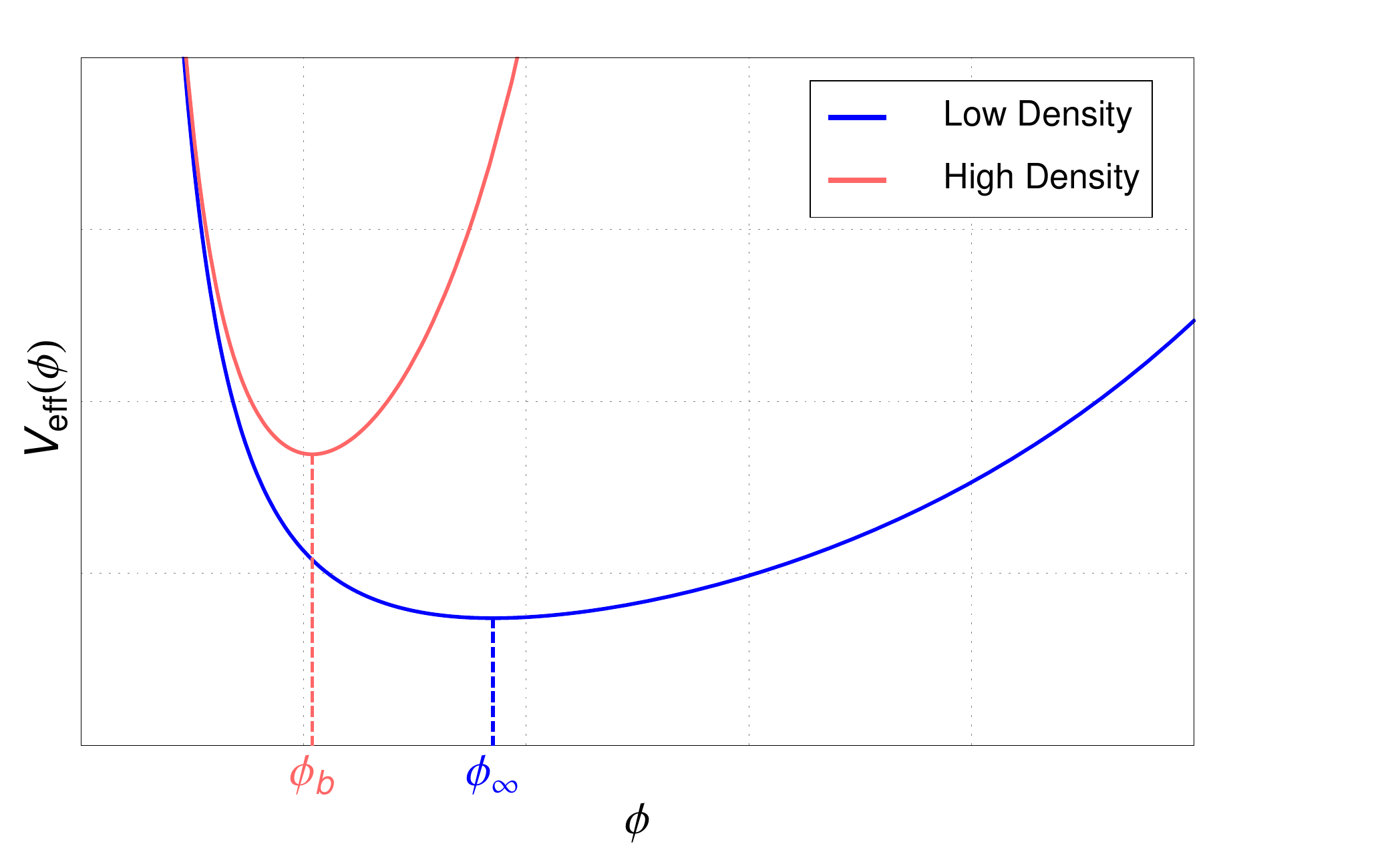}
\end{center}
\caption{Representation of the effective potential for large and small value of the matter density $\t\rho$. This illustrates that, as $\t\rho$ decreases, the effective potential becomes wider and the minimum shifts to higher value. This situation can represent the effective potential inside a body (star or planet) in red and outside the body in blue.} 
\label{figEffPot}
\end{figure}

As already stated in the previous section, in the non-relativistic limit, the evolution of $\phi$ is governed by the effective potential $V_{\rm eff}(\phi,\t\rho)$ which depends explicitly on the local matter density. This potential is illustrated on Fig.~\ref{figEffPot} where the blue line represent the effective potential outside the body (the cosmological density $\t\rho_\infty$ is much smaller than the body density $\t\rho_b$) and the red line represents the potential inside a body (star or planet).
The minimum of the potential is noted $\phi_{min}$ and the mass of small fluctuations around $\phi_{min}$ is obtained by evaluating the second derivative of the potential
\begin{eqnarray}
	m^2&=&\left.\frac{\partial^2V_{\rm eff}	}{\partial \phi^2}\right|_{\phi=\phi_{min}}\nonumber\\
	&=&\frac{1}{m_p^2}\left(\frac{d^2V(\phi_{min})}{d\phi^2}+4A^4(\phi_{min})k^2\t\rho\right)   \cdot
\end{eqnarray}
As can be seen on Fig.~\ref{figEffPot}, larger value of $\t\rho$ corresponds to smaller value of $\phi_{min}$ and to larger value of $m$ (the potential becomes more narrow). In other word, the denser the environment, the more massive the chameleon. In what follow, indices $b$ refers to quantity evaluated at the minimum of the effective potential inside the body and $\infty$ refers to quantity evaluated at the minimum of the effective potential outside the body.

\begin{figure}[htb]
\begin{center}
\includegraphics[width=0.5\textwidth]{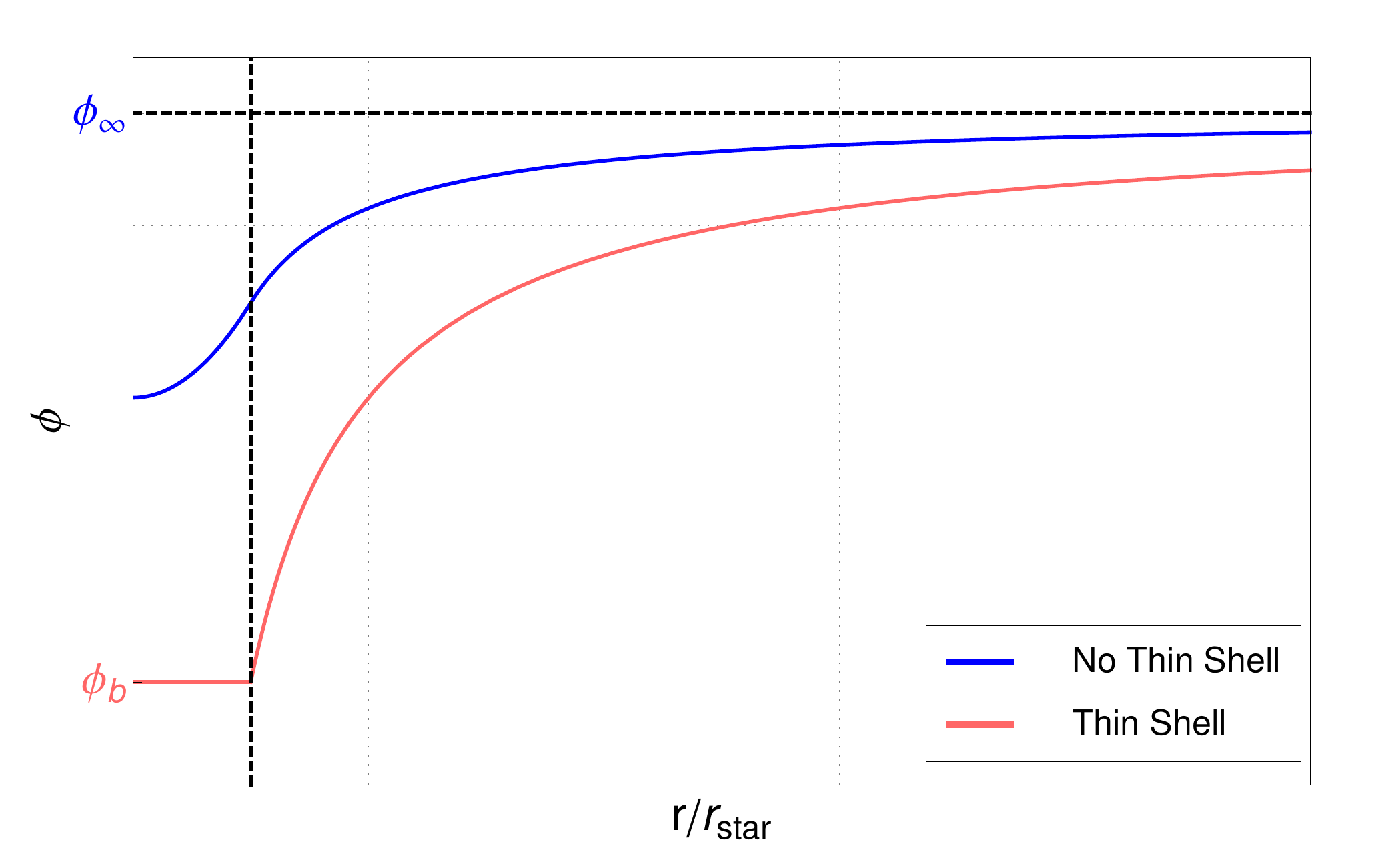}
\end{center}
\caption{Representation of the evolution of the scalar field as a function of the radius coordinate in the two regimes. In red (ligth): in the thin-shell regime, we see that the field is frozen in the minimum of the potential inside the star. In blue (dark): the thick-shell regime, the field does not reach the minimum of the effective potential inside the star.} 
\label{figThinShell}
\end{figure}

In chameleon model, it is assumed that the scalar field reaches its minimum far away from the body ($\phi(r=\infty)=\phi_\infty$) as we discussed around (\ref{phi_inf}). On the other side, the field does not necessary reach its minimum inside the body. Two regimes are possible:
\begin{itemize}
	\item the field does effectively reach the minimum of the potential inside the body ($\phi(r=0)\approx\phi_b$) and is frozen in this minimum in a large part of the body. This regime is called the \emph{thin-shell} regime~\cite{khoury:2004uq,khoury:2004fk} because the field evolves only in a small shell around the body's surface. If we note by $R_b$ the body radius and by $R_r$ the radius where the field begins to move, the condition to have a thin-shell regime is given by 
	\begin{equation}
		 \frac{\Delta R_b}{R_b}=\frac{R_b-R_r}{R_b}<1
	\end{equation}
	 Following Khoury and Weltman~\cite{khoury:2004uq,khoury:2004fk}, we can write this condition as follow
\begin{equation}
		\frac{\Delta R_b}{R_b}\approx \varepsilon=\frac{\phi_{\infty}-\phi_b}{\frac{3}{4\pi}k\Phi_b}<1
\end{equation}
where $\varepsilon$ is called the thin-shell parameter \footnote{The definition of the thin-shell parameter $\varepsilon$ is different from the definition given in Khoury and Weltman~\cite{khoury:2004fk,khoury:2004uq} because the starting action is not the same. In particular, our scalar field $\phi$ is related to their scalar field $\phi_{KW}$ by the relation $\phi=\frac{\phi_{KW}}{m_p}$ and as a consequence, our scalar coupling $k$ is related to their scalar coupling $\beta$ by $k=\sqrt{8\pi}\beta$. With these relations, the thin-shell parameters defined in Khoury and Weltman~\cite{khoury:2004fk,khoury:2004uq} $\varepsilon=\frac{\phi_{KW\infty} - \phi_{KWb}}{6\beta M_p \Phi_b}$ becomes $\frac{\phi_{\infty}-\phi_b}{\frac{3}{4\pi}k\Phi_b}$. }                                       
and $\Phi_b$ is the Newtonian    potential at the body's surface.

\item the field does not reach the minimum of the potential inside the body and the field is not frozen at all inside the body. This regime is called the \emph{thick-shell} regime.
\end{itemize}
Fig.~\ref{figThinShell} shows qualitatively the difference between the two regimes. In the thin-shell regime, the field is frozen around $\phi_b$ inside the body while this minimum of the effective potential is not reached in the thick-shell case. 
   
\subsection{Field profile and fine-tuning} \label{sec:num}
The field profile $\phi(r)$ has been derived analytically by Khoury and Weltman~\cite{khoury:2004fk}. A more precise solution has been derived by Waterhouse~\cite{waterhouse:2006kx} and by Tamaki and Tsujikawa~\cite{tamaki:2008ve}. They introduced a third regime: the \emph{no-shell} regime which essentially corresponds to the thin-shell regime in the limit where the thin-shell vanishes. These solutions have been derived by considering a lot of hypothesis:
\begin{itemize}
	\item   the solution has been derived using a Minkowski background space-time. 
	\item outside the body, the derivative of the effective potential is linearized
	\item in the thin-shell regime, inside the body the derivative of the effective potential is also linearized
	\item $m_\infty R_b<<1$
	\item the factor $A^4(\phi)$ is always approximated to 1 (which means that $k\phi << 1$)
\end{itemize}
The validity of the previous hypothesis have never been checked afterwards by comparing the analytical solution with the numerical solution of the full relativistic system of equations (\ref{fieldSun}).
Finally, Tsujikawa et al.~\cite{tsujikawa:2009qf} derived the field profile considering a perturbation coming from the gravitational background. 

The exterior solution derived with the previous hypothesis is given by
\begin{equation}   \label{phiprof}
	\phi(r)=\phi_\infty - \frac{k_{\rm eff}}{4\pi}\frac{GM_b}{rc^2}e^{-m_\infty(r-R_b)}
\end{equation}                                                                     
where $k_{\rm eff}$ depends on the thin-shell parameter $\varepsilon$
\begin{equation}\label{epsilon}
\varepsilon=\frac{\phi_{\infty}-\phi_b}{\frac{3}{4\pi}k\Phi_b}\cdot
\end{equation}
Two regimes are identified depending on the value of $\varepsilon$:
\begin{itemize}
	\item if $\varepsilon<\frac{1}{2}+\frac{1}{(m_bR_b)^2}$. This case corresponds to the thin-shell regime and the effective coupling constant is given by 
\begin{subequations}\label{keff}
\begin{eqnarray}
k_{\rm eff}&=&3k\left[\frac{\Delta R_b}{R_b}+\frac{1}{m_bR_b}\right.\label{keff1}\\
&-&\left.(\frac{\Delta R_b}{R_b})^2-\frac{2}{m_bR_b}\frac{\Delta R_b}{R_b}-\frac{1}{(m_bR_b)^2}\right]\approx 3k\varepsilon\cdot \nonumber
\end{eqnarray} 
	\item if  $\varepsilon>\frac{1}{2}+\frac{1}{(m_bR_b)^2}$. This case corresponds to the thick-shell regime and the effective coupling constant is given by 
\begin{equation}\label{keff2}
	          k_{\rm eff}=k\cdot
\end{equation}                  
\end{subequations}
\end{itemize}
The analytical expression for the interior solution can be found for example in Tamaki and Tsujikawa~\cite{tamaki:2008ve} but will not be needed in our case since we will be only interested in the exterior solution (which represents the solution around the Sun).

\begin{figure}[htb]
\begin{center}
\includegraphics[width=0.45\textwidth]{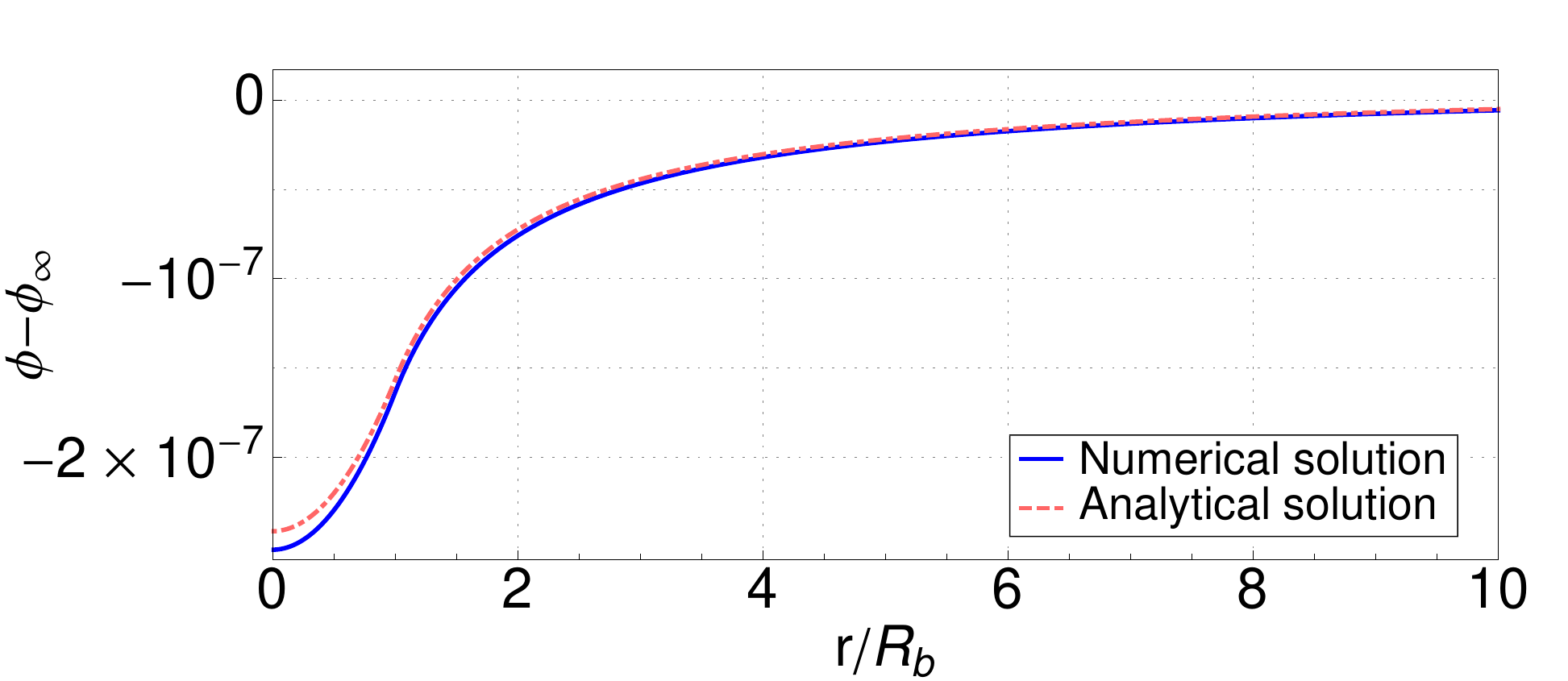}
\includegraphics[width=0.45\textwidth]{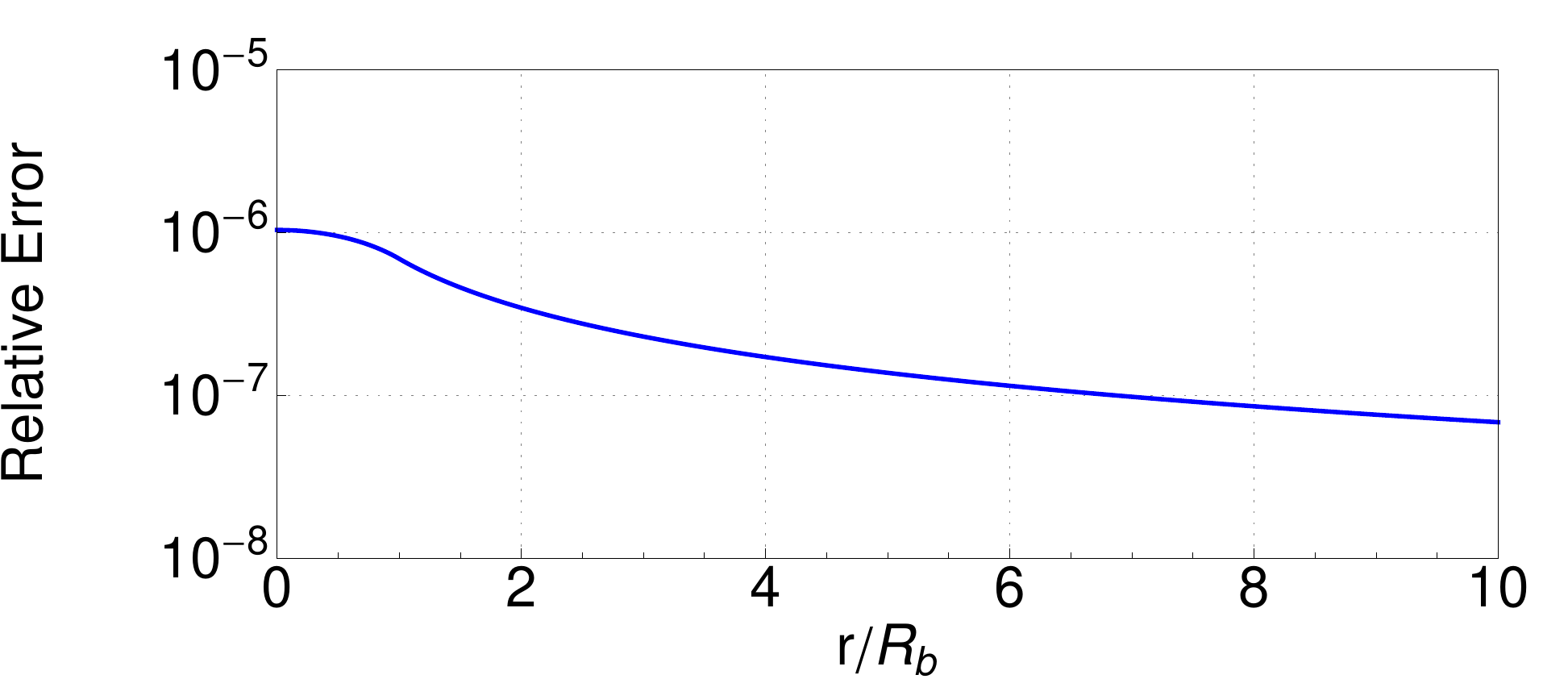}
\end{center}
\caption{Representation of the evolution of the scalar field around the Sun for a model characterized by the parameters: $\Lambda=10^{-9}\ GeV$, $\alpha=0.5$, $k=1$ and $\t\rho_{\infty}=10^{-27}kg/m^3$. With these parameters, we obtain a thick-shell around the sun. On top: comparison of the field profile around the neighborhood of the Sun obtained numerically by solving the full BVP problem (\ref{fieldSun}) and by using the analytical solution. On bottom: relative difference between the numerical solution and the analytical approximation. One can see that the analytical solution is very good far from the sun but is less efficient in the Sun.} 
\label{figThickNum}
\end{figure}

Fig.~\ref{figThickNum} represents a comparison between the analytical solution and the numerical solution of the full BVP problem (\ref{fieldSun}) for a situation corresponding to the sun and with a cosmological density of $10^{-27}kg/m^3$ for a model giving a thick-shell. We can see that the analytical solution is very good far from the sun but this agreement is less good inside the sun. This is mainly due to the fact that the analytical solution is obtained by neglecting the $A^4(\phi)$ term in the equation (\ref{psi_}). Since, for the determination of the PPN parameters, we only need the outside solution, the analysis of Fig.~\ref{figThickNum} shows that the analytical solution can be used in the case of thick-shell.  
            
The case of a thin-shell profile around the Sun is more delicate to treat from a numerical point of view. Indeed, it is no longer possible to solve numerically the full BVP. This is mainly due to the fact that $\t\rho_b/\t\rho_\infty \sim 10^{30}$ which means that, in order to have a model exhibiting a thin-shell profile, the effective mass of the scalar field inside the body ($m_b$) needs to be huge (as can be inferred from the relation (83) of Tsujikawa et al.~\cite{tsujikawa:2009qf}). As a result of this huge effective mass $m_b$, variations of the scalar field inside the body are extremely small and very often smaller than the epsilon machine. This leads to problems very delicate to treat numerically. The thin-shell profile obtained by solving the full BVP was obtained by Babichev and Langlois~\cite{babichev:2010bh} but in the case where the ratio of density was of the order of $10^{-2}-10^{-4}$. 

Instead of solving the full BVP, which is very hard to do, we treat a simplified BVP constituted of the scalar field only. This means that we approximate the space-time background by a Minkowski space-time (this is also one of the hypothesis done in order to get the analytical solution). In this case, the one second order differential equations with boundary condition ($\phi'(r=0)=0$ and $\phi(r=\infty)=\phi_\infty$) is solved by a shooting method implemented with a stiff integrator in quadruple precision. With this method, it is possible to get thin-shell profile around the Sun.  

\begin{figure}[htb]
\begin{center}
\includegraphics[width=0.45\textwidth]{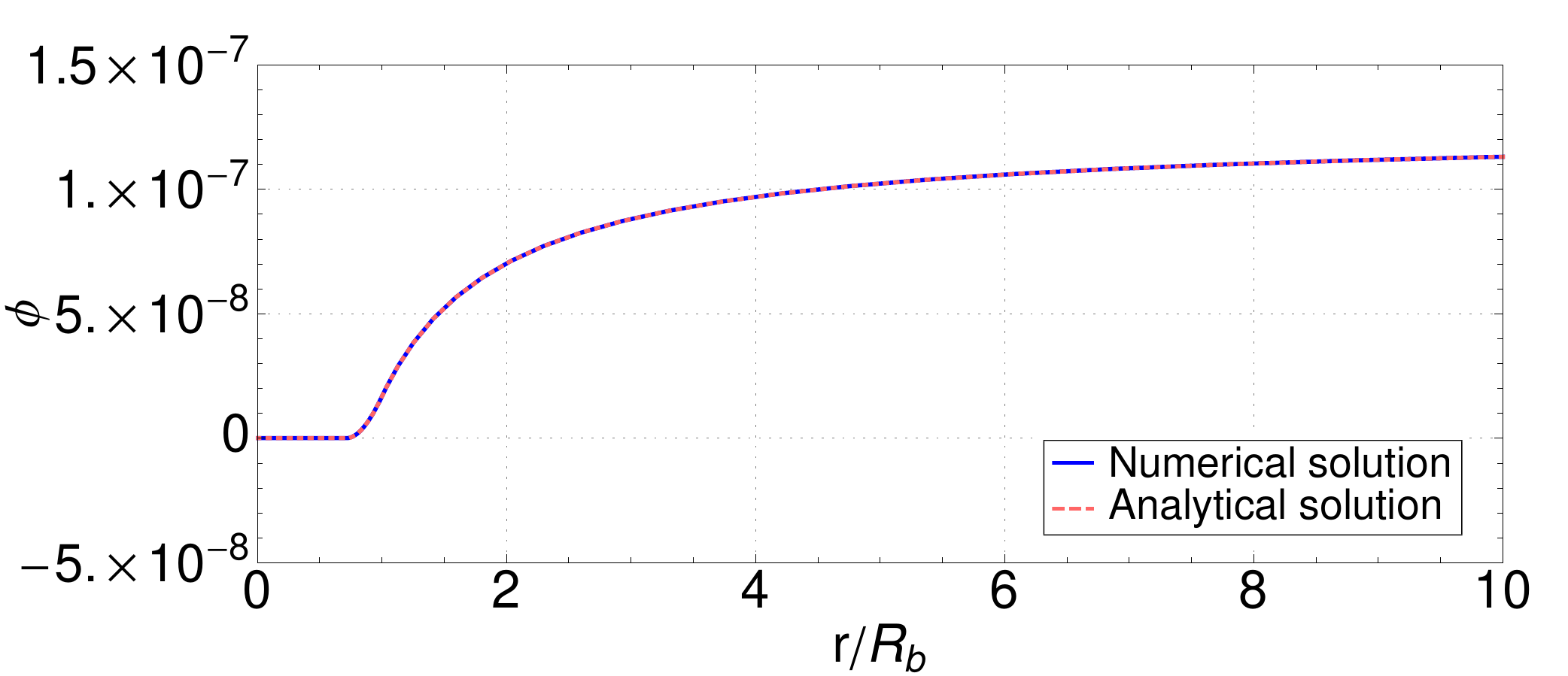}
\includegraphics[width=0.45\textwidth]{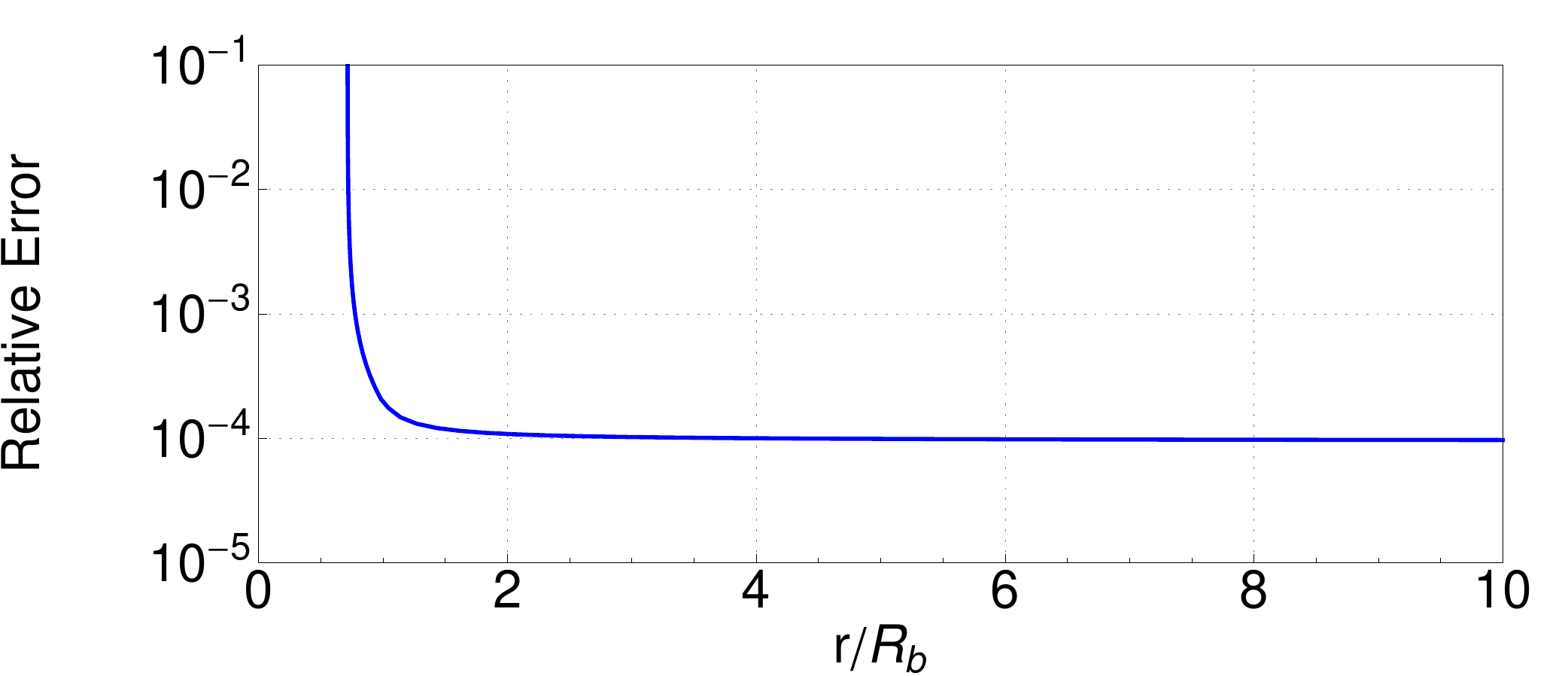}
\end{center}
\caption{Representation of the evolution of the scalar field around the Sun for a model characterized by the parameters: $\Lambda=10^{-2} \ GeV$, $\alpha=3.3$, $k=1$ and $\t\rho_{\infty}=10^{-27}kg/m^3$. With these parameters, we obtain a thin-shell around the sun. On top: comparison of the field profile around the neighborhood of the Sun obtained numerically by solving the BVP problem related to the scalar field and by using the analytical solution. On bottom: relative difference between the numerical solution and the analytical approximation. One can see that the analytical solution is very good far from the sun but is less efficient in the Sun.} 
\label{figThinNum}
\end{figure}

Fig.~\ref{figThinNum} represents a comparison between the analytical solution and the numerical solution of the simplified BVP for a situation corresponding to a spherical sun and with a cosmological density of $10^{-27}kg/m^3$ for a model giving a thin-shell. We can see that the analytical solution is very good far from the sun but this agreement is less good inside the sun. Once again, for the determination of the PPN parameters, we only need the outside solution. The analysis of Fig.~\ref{figThinNum} shows that the analytical solution can be used in the case of thin-shell.      

\subsection{Post-Newtonian parameters} \label{sec:ppn}
In order to constrain the parameters characterizing the theory with solar system experiments, it is useful to derive the post-newtonian parameters. As indicated in relation (\ref{ppn1}), the post-newtonian parameters for a tensor-scalar theory without potential is given by
$$
	\t\gamma-1 =-2\frac{k^2}{4\pi+k^2},\qquad \t\beta-1=0\cdot
$$
As shown in the previous section, when a potential is added to the theory, the chameleon effect can reduced the coupling constant $k$ to an effective coupling constant $k_{\rm eff}$ as indicated in (\ref{keff}).  Considering the chameleon theory as a Brans-Dicke theory with an effective coupling constant $k_{\rm eff}$ as it was done in the original Chameleon papers~\cite{khoury:2004fk,khoury:2004uq} is not correct. Instead, one has to compute more carefully the PPN parameters following the procedure described in Damour and Esposito-Far\`ese~\cite{damour:1992ys}. This procedure is used in Faulkner et al.~\cite{faulkner:2007kx} but they used a definition of the Post-Newtonian Parameters in Schwarzschild coordinates which can be misleading since the original PN parameters are defined in Standard Post Newtonian gauge~\cite{will:1993fk} where the coordinates are isotropic. More precisely, the PN parameters are defined with the Jordan-frame metric in isotropic coordinates as
\begin{eqnarray}
	d\t s^2&\approx&-\left(	1-2\frac{\t G \t M_b}{c^2 \t \chi}+2\t \beta\left(\frac{\t G \t M_b}{c^2 \t \chi}\right)^2\right)d\t t^2 \nonumber\\
	&&+ \left(1+2\t\gamma \frac{\t G \t M_b}{c^2 \t \chi} \right)\left(d\t\chi^2+\t\chi^2d\t\Omega^2\right)  \label{ppnmetric}
\end{eqnarray}                                                                                        
where $\t\chi$ is the Jordan frame isotropic radial coordinate. In all this section, we consider weak gravitational field and therefore we use a Post-Newtonian expansion. In this context, the symbol $\approx$ emphasizes the fact that the equality is valid only to first Post-Newtonian order.
Instead of using isotropic coordinates, one can perform the transformation to express the space-time metric in Schwarzschild coordinates (this transformation for constants PN parameters is derived in~\cite{weinberg:1972vn}) using the transformation
\begin{equation}
	\t r\approx\t\chi\left(1+\gamma \frac{\t G\t M_b}{c^2\t\chi}\right)\cdot
\end{equation}                                                  
A simple calculation gives
\begin{eqnarray}
	d\t s^2 &\approx  &- \left(1-2\frac{\t G \t M_b}{c^2 \t r}+2(\t\beta-\t\gamma)\left(\frac{\t G\t M_b}{c^2\t r}\right)^2\right)d\t t^2\nonumber \\
	&&+\left(1+2\t\gamma \frac{\t G \t M_b}{c^2\t r}-2\t\gamma'\frac{\t G \t M_b}{c^2}\right)d\t r^2+\t r^2d\t\Omega^2,
\end{eqnarray}                                                                                        
where $\t\gamma'=d\t\gamma/d\t r$ is the derivative of the PN parameter $\t\gamma$ with respect to the Schwarzschild radial coordinates. This simple calculation shows that when we are dealing with space-time dependent PN parameters, one has to be very careful with the gauge in which the PN parameters are defined since its spatial derivative can appear. For example, in Faulkner et al.~\cite{faulkner:2007kx}, they used Schwarzschild coordinates to identify a space dependent $\t\gamma$ but without any term related to $\t\gamma'$. This means that the expression they derive will differ from the one defined in isotropic coordinates which is the definition of $\t\gamma$ given in Will~\cite{will:1993fk}.  

In the following, we will follow the procedure described in Damour and Esposito-Far\`ese~\cite{damour:1992ys} being very careful to the gauge used. In the following, $\chi$ will refer to isotropic radial coordinate and $r$ to Schwarzschild radial coordinate and the $\tilde {}$ always refers to Jordan frame quantities. We start from the spherically symmetric metric in the Einstein frame in the weak field limit (in isotropic coordinates):
\begin{eqnarray}
	ds^2&\approx&-\left(1-2\frac{M_b}{m_p^2\chi}+2\left(\frac{M_b}{m_p^2\chi}\right)^2\right)dt^2 \nonumber\\
	&&+   \left(1+2\frac{M_b}{m_p^2\chi}\right)(d\chi^2+\chi^2d\Omega^2).
\end{eqnarray}                                                                                                                          
Then, we can perform the conformal transformation to derive the Jordan frame metric
\begin{equation}
	d\t s^2=A^2(\phi)ds^2\cdot
\end{equation}                
To transform the last expression into the form (\ref{ppnmetric}), one uses the field profile (\ref{phiprof}) as function of the Schwarzschild radial coordinates and the transformation between Schwarzschild and isotropic coordinates ($r\approx\chi+M_b/m_p^2$) to express the field profile as function of $\chi$. Finally, using the expansion $A^2(\phi)=A^2(\phi_\infty)e^{2k(\phi-\phi_\infty)}=A_\infty^2(1+2k(\phi-\phi_\infty)+2k^2(\phi-\phi_\infty)^2)$  and the relations between Jordan and Einstein frame quantities~\cite{damour:1992ys} ($\t x^\mu=A(\phi_\infty)x^\mu=A_\infty x^\mu$ and $m_b=A_\infty\t m_b$), one obtains the metric (\ref{ppnmetric}) with 
\begin{subequations}
\begin{eqnarray}
	\t G & =  & \frac{A(\phi_\infty)^2}{m_p^2}\left(1+\frac{kk_{\rm eff}}{4\pi}\right)e^{-m_\infty(\chi-\chi_b)} \\
	\t\gamma &=& \frac{4\pi-kk_{\rm eff}e^{-m_\infty(\chi-\chi_b)}}{4\pi+kk_{\rm eff}e^{-m_\infty(\chi-\chi_b)}}\\
	\t\beta&=&0\cdot
\end{eqnarray}     
\end{subequations}
In practice, $m_\infty$ is tiny and the exponential decay can be neglected leading to the expression
\begin{equation}
\t\gamma = \frac{4\pi-kk_{\rm eff}}{4\pi+kk_{\rm eff}}\label{gammacham}
\end{equation}
This approach gives the PN parameters as defined by Will~\cite{will:1993fk} and avoid any confusion due to the choice of coordinates to define the $\t\gamma$ parameter. However, in a lot of papers~\cite{de-felice:2010uq,faulkner:2007kx,capozziello:2008vn} this derivation is done using Schwarzschild coordinates without the term linked to the derivative $\t\gamma'$. For this reason, we expect differences in the expression of the PN parameter $\t\gamma$ due to the difference of gauge used. This difference is present in relation (5.50) of De Felice and Tsujikawa~\cite{de-felice:2010uq} where an additional term is present in the denominator of $\t\gamma$. Nevertheless in~\cite{faulkner:2007kx,capozziello:2008vn}  $m_\infty$ is supposed to be very small and they neglect terms that are in fact closely related to the $\t\gamma'$ contribution and they find a result similar to (\ref{gammacham}) (with some differences due to the choice of the starting action, see footnote~\ref{footphi}). Finally considering the theory equivalent to Brans-Dicke (which is done in Khoury and Weltman~\cite{khoury:2004fk,khoury:2004uq}) with a coupling constant $k_{\rm eff}$ leads to a $\t \gamma=(4\pi-k_{\rm eff}^2)/(4\pi+k_{\rm eff}^2))$ which is quite different than expression (\ref{gammacham}) and which leads to a misevaluation of the PN parameters.    
         
The thin-shell effect has the property to hide the deviations from GR on solar system scales. Indeed, in the case of a thin-shell profile, the deviation from GR is quantified by the quantity
\begin{equation}
	\t\gamma-1=-2\frac{kk_{\rm eff}}{4\pi+kk_{\rm eff}}\approx  -6\frac{\varepsilon k^2}{4\pi+3\varepsilon k^2}\cdot
\end{equation}   
This shows that big values of the coupling constant $k$ can be compatible with solar system experiments such as the measurement of the Shapiro delay by Cassini (\ref{cassini}) if the thin-shell parameter $\varepsilon$ is small enough while big values of $k$ are completely excluded in Brans-Dicke theory. Using (\ref{epsilon}), the last expression becomes
\begin{equation}
	\t\gamma-1=-2\frac{k(\phi_\infty-\phi_b)}{\Phi_b+k(\phi_\infty-\phi_b)}\cdot
\end{equation}                                                                  
It is possible to push the analytical development further by considering that $\phi_\infty >> \phi_b$ (which is justified since it is shown in Tsujikawa et al.~\cite{tsujikawa:2009qf} that $\phi_\infty/\phi_b=(\t\rho_b/\t\rho_\infty)^{1/(\alpha+1)}\sim 10^{30/(\alpha+1)}$ if we use the cosmological and the sun density). Isolating $k\phi_\infty$ from the last relation
\begin{equation}
	k\phi_\infty=-\frac{\Phi_b(\t\gamma-1)}{2+(\t\gamma-1)},
\end{equation}
and using the Cassini constraint (\ref{cassini}) and the value of the Newton potential at sun radius $\Phi_b=2.12 \ 10^{-6}$, we get some boundary value $k\phi_\infty \leq 2.12 \ 10^{-12}$. The value of $\phi_\infty$ is given by the value of the field minimizing the effective potential $V_{\rm eff}$ and is given by relation (\ref{phimin}). For very low value of $\phi_\infty$ (which is the case here), we can use the Taylor expansion of the W-Lambert function $W_L(x)\approx x$ and we find the constraint
\begin{equation} \label{cassthinshell}
	k\left(\frac{\alpha\Lambda^{4+\alpha}}{m_p^\alpha k\t\rho_\infty}\right)^{\frac{1}{\alpha+1}}\leq 2.12 \ 10^{-12}\cdot
\end{equation}  

Let us summarize the discussion about Post-Newtonian parameters. The $\t\gamma$ post-newtonian parameters has been computed with great care of the gauge used. The $\t\gamma$ PN parameter is given by expression (\ref{gammacham}) where $k_{\rm eff}$ depends on the thin-shell parameters $\varepsilon$. If $\varepsilon>\frac{1}{2}+\frac{1}{(m_bR_b)^2}$, we have a thick-shell and the theory is equivalent to Brans-Dicke and the Cassini constraint gives the usual constraint on the coupling constant $k^2\leq 10^{-4}$. If $\varepsilon<\frac{1}{2}+\frac{1}{(m_bR_b)^2}$, we have a thin-shell and the Cassini constraint is given by the constraint (\ref{cassthinshell}).

\begin{figure}[htb]           
\begin{center}
\includegraphics[width=0.45\textwidth]{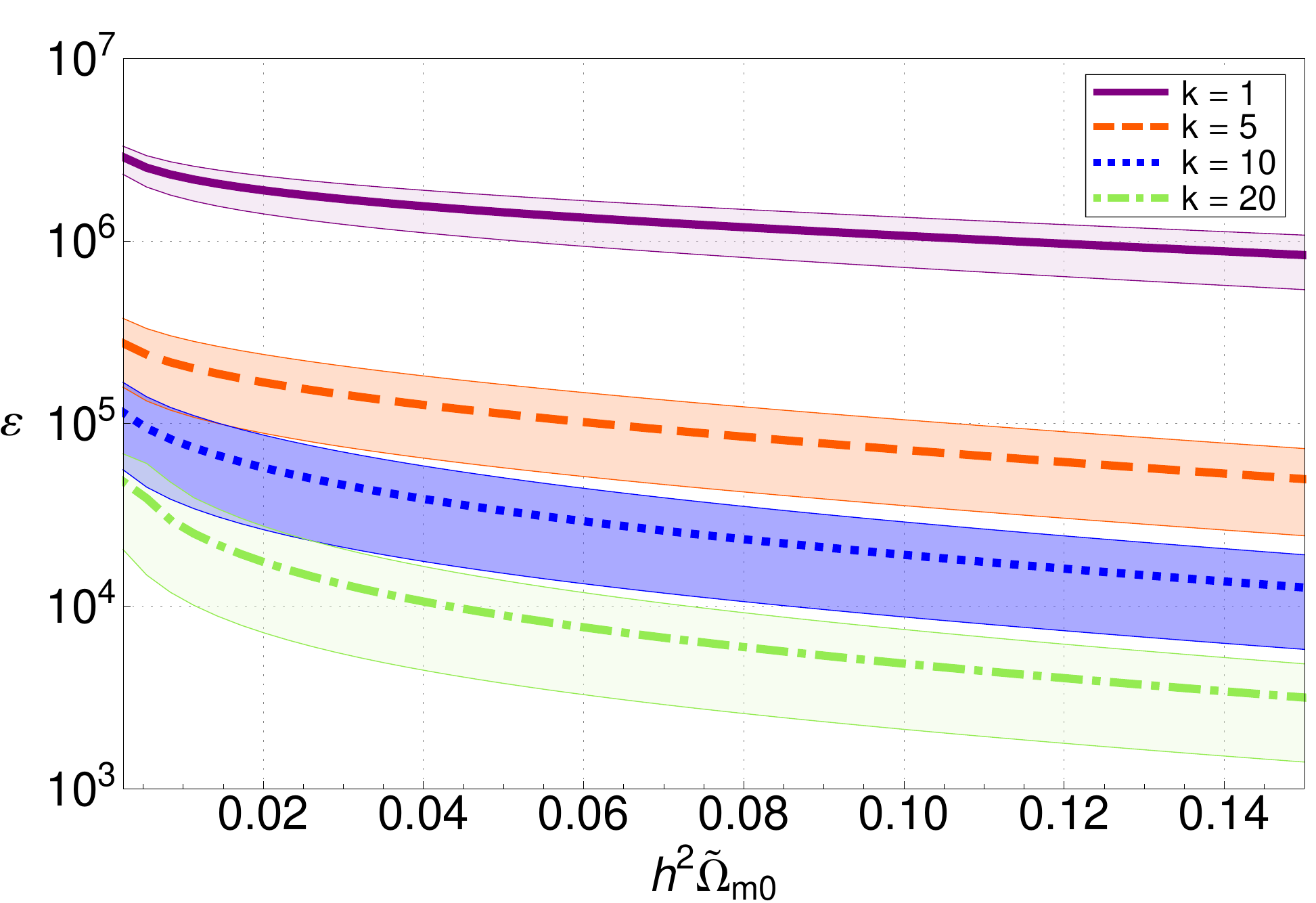}
\end{center}
\caption{Representation of the thin-shell parameters $\varepsilon$. The different lines represent different values of the coupling constant and the filled areas represent different values of $\alpha$ between 0.5 and 3.} 
\label{figEpsilon}
\end{figure}

For the models within the 95\% confidence region of the supernovae Ia likelihood analysis (see Section~\ref{sec:likelihood}), we compute the thin-shell parameter $\varepsilon$ given by relation (\ref{epsilon}). Fig.~\ref{figEpsilon} shows the evolution of $\varepsilon$ for model considered in Section~\ref{sec:likelihood}. We can see that all these models present a huge value of the thin-shell parameter. This means that no thin-shell mechanism is playing any role for all these models and they are all equivalent to Brans-Dicke theory. Therefore, for models explaining cosmic acceleration, no chameleon effect is present on solar system scale. Since the theory is equivalent to Brans-Dicke, the constraint on $\t\gamma$ gives $k^2<10^{-4}$.

\begin{figure}[htb]
\begin{center}
\includegraphics[width=0.45\textwidth]{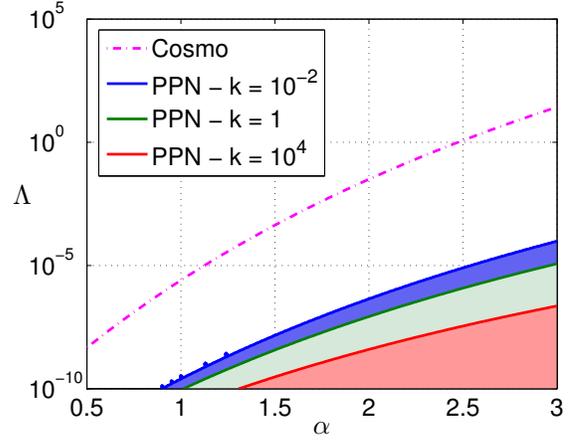}
\end{center}
\caption{Representation of couples $(\alpha,\Lambda)$ admissible with the Cassini constraint (\ref{cassini}) for different values of the coupling constant $k$ and representation of the couples $(\alpha,\Lambda)$ needed to explain cosmological observations. The density at infinity of the solar system $\t\rho(r=\infty)=10^{-27} kg/m^3$ which corresponds to the cosmological matter density.} 
\label{figCombConst1}
\end{figure}

Another way to see the incompatibility between cosmological observations and solar system observations is to take the $(\alpha,\Lambda)$ plane already presented on Fig.~\ref{figLambda}. With relations (\ref{gammacham}) and  (\ref{keff}), it is possible to compute the $\t\gamma$ post-newtonian parameters for different models characterized by the three parameters $\Lambda, \alpha$ and $k$ (considering a cosmological observed matter density of $10^{-27} \ kg/m^3$). Fig.~\ref{figCombConst1} represents the area in the parameter space $(\alpha,\Lambda)$ that are compatible with $\t\gamma$ constraint (\ref{cassini}) for different values of the coupling constant. Several remarks need to be done. First of all, we see that high coupling constants are not completely rejected by solar system experiments. For example, $k=1$ or $k=10^4$ are completely rejected by considering massless tensor-scalar theory while it can be seen that these coupling constants can be admissible with respect to post-newtonian test of gravity in a certain domain of the space parameters $(\alpha,\Lambda)$. In particular, for low values of $\Lambda$ and high values of $\alpha$, the chameleon effect is strong enough to reduce the scalar charge $k_{\rm eff}$ so that the $\t\gamma$ parameter (\ref{gammacham}) satisfies the Cassini constraint. This was the breakthrough of the original papers ~\cite{khoury:2004fk,khoury:2004uq}. Nevertheless, we can see on Fig.~\ref{figCombConst1} that it is not possible to explain cosmic acceleration and to pass the solar system experiments at the same time (at the exception of the small coupling constants $k$ which is also the case for Brans-Dicke theory). Indeed, there is no intersection between the curves representing the relation between $\Lambda$ and $\alpha$ imposed by the cosmological observations and the domain allowed by the PPN constraints. 

\begin{figure}[htb]
\begin{center}
\includegraphics[width=0.45\textwidth]{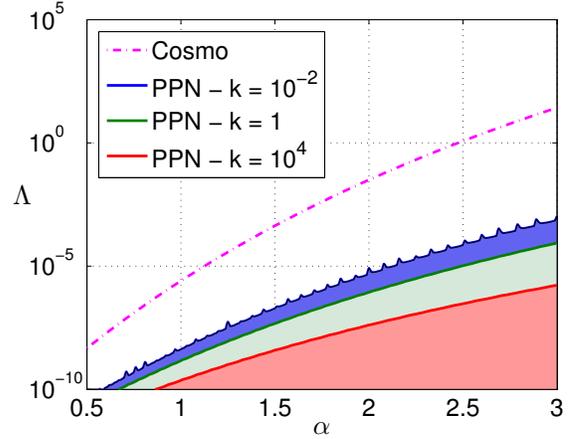}
\end{center}
\caption{Representation of couples $(\alpha,\Lambda)$ admissible with the Cassini constraint (\ref{cassini}) for different values of the coupling constant $k$ and representation of the couples $(\alpha,\Lambda)$ needed to explain cosmological observations. The density at infinity of the solar system $\t\rho(r=\infty)= \ 10^{-21} kg/m^3$ which roughly corresponds to the galactic matter density.} 
\label{figCombConst2}
\end{figure}

In the calculation of the PPN parameters in Fig.~\ref{figCombConst1}, the density at infinity of the solar system is assumed to be the cosmological density ($\t\rho(r=\infty)=\t\rho_0$). This situation is a simplified situation. In reality, the situation is much more complex since the solar system is not plunged into a cosmological fluid but is a part of a galaxy which itself is a part of a galaxy cluster and so one\dots As a consequence, the scalar field will have an evolution at all the different scales from solar system scales to cosmological scales (passing by the galactic scale and the galaxy cluster scale and so one\dots). All this modeling is quite complex and beyond the scope of this paper. Nevertheless to have some intuition, we compute the PPN parameter by assuming that the density at infinity of the solar system is given by the galactic density at a distance of the solar system. We assume a galactic density of $\t\rho_{gal}\approx 10^{-21} kg/m^3$ at solar system distance from the galactic center. Fig.~\ref{figCombConst2} represents the admissible domain in the $(\alpha,\Lambda)$ plane supposing that the density at infinity is given by the galactic density  $\t\rho(r=\infty)=\t\rho_{gal}$. We can see that the situation is better than considering the cosmological density. The area of the region satisfying the PPN constraint are bigger but there is still no intersection between the cosmological relation and the solar system constraint.

The results of this analysis is that, on solar system scales, the chameleon effect can allow high coupling constant $k$ that are normally excluded by Brans-Dicke analysis at the condition that parameters $\alpha$ and $\Lambda$ are in the admissible domain (which mean for high $\alpha$ and small $\Lambda$). But we have shown that there are no intersection between models explaining cosmological expansion and models authorized by solar system experiments.

\section{Conclusion}
In this paper we have studied massive tensor-scalar theories at two different scales: on cosmological scales and on solar system scales. We have presented combined constraints of the so-called chameleon field introduced by Khoury and Weltman~\cite{khoury:2004fk,khoury:2004uq}. In this communication we have only focused on the original model which is characterized by an exponential coupling function $A(\phi)=e^{k\phi}$ and on a Ratra-Peebles~\cite{ratra:1988vn} run-away potential. We have also derived a unambiguous definition of observables like luminous distance and $\t\gamma$ Post-Newtonian parameter, leading to different predictions than in literature.
      \\

On cosmological scales, we have derived the cosmological equations of evolution by using the Jordan frame density parameter as source term in the Einstein equations. Using the latest Supernovae Ia data~\cite{kowalski:2008zr}, we have derived confidence regions for the coupling constant $k$, for the parameters characterizing the potential and for the cosmological matter density today $\t\Omega_{m0}$. Considering models within this confidence region, we have derived a relationship between the two parameters characterizing the Ratra-Peebles potential. The relationship found (\ref{lambdaAlpha}) is exactly the same as the one obtained in quintessence model (characterized by a vanishing coupling constant $k=0$)~\cite{schimd:2007vn}.  A detailed analysis of the cosmological evolution has been presented. In particular, the contribution of the scalar field has been parametrized by an effective fluid of dark energy in GR. We have presented the evolution of the effective equation of state parameter $\t\omega_{DE}$ and we have shown that the derivative of this parameter can become positive which can lead to a test to distinguish tensor-scalar theories from quintessence scenarios. Moreover, we have shown that for high coupling constants, the cosmic acceleration is mainly produced by the non-minimal coupling and not by the scalar self interaction. Finally, we identified two dynamical regimes that can explain the cosmic acceleration. This two regimes are characterized by two minima in the $\chi^2$ curves as shown in Fig.~\ref{figTrustAge}.
\\

In the second part of the article, we have studied the solar system configuration modeled by a spherical and static central body (the Sun). We have computed numerically the field profile solving the full boundary value problem composed by the Einstein field equations and we have compared these numerical simulations to analytical solutions that can be found in the chameleon literature. In particular, we have shown that the analytical solutions are very good outside the central body but are not completely satisfactory inside the body. These simulations have confirmed that the thin-shell mechanism can appear in the surrounding of the Sun. Using analytical  expression outside the central body, we have derived the post-newtonian parameters. In this derivation we have taken a great care to the gauge used to define PN parameters comparing our approach with existing expressions in chameleon literature. In the case of a thin-shell situation we have confirmed that the scalar charge of the central body can drastically be reduced by the relation $k_{\rm eff}=3\varepsilon k$ where $\varepsilon$ is the thin-shell parameter. Contrary to what is claimed in Khoury and Weltman~\cite{khoury:2004fk,khoury:2004uq}, we have shown that the theory is not equivalent to Brans-Dicke theory with an effective coupling constant $k_{\rm eff}$. As a consequence, the $\t\gamma$ constraint is different from what was derived in the original chameleons papers. Nevertheless, we have confirmed that in the case of a thin-shell situation, the reduction of the scalar charge implies a reduction of the deviation of the post-newtonian parameter $\t\gamma$ from its GR value. This confirms the fact that high values of the coupling constant $k$ can eventually pass solar system constraints while these coupling constants were previously excluded by considering traditional tensor-scalar theories. 

For models within the 2$\sigma$ cosmological confidence region we have computed the thin-shell parameters $\varepsilon$ and we have found that this parameter is always very huge compared to the unity ($\varepsilon >> 1$). This means that no thin-shell effect can be invoked on solar system scales for models fitting Supernovae Ia data and as a consequence no reduction of the scalar charge can be observed. The theory is therefore equivalent to Brans-Dicke and the Post-Newtonian constraint implies that only small coupling constants are admissible. Another presentation of the combined constraints have been presented in the form of admissible region in the $(\alpha,\Lambda)$ plane (where $\Lambda$ and $\alpha$ are the parameters characterizing the Ratra-Peebles potential). In this plane, we have shown that there is no intersection between cosmological admissible domain and solar system admissible domain. 
\\

The main conclusion of this work is that the chameleon field (if one consider the initial model by Khoury and Weltman~\cite{khoury:2004fk,khoury:2004uq}) can not explain cosmological accelerated expansion and pass solar system constraints  at the same time. On one hand, we have shown that there are models fitting Supernovae Ia data relying most on non minimal coupling than on self interaction potential. On the other hand, there are models satisfying solar system constraints even for high coupling constant (which are rejected if no chameleon mechanism is present). But there are no models satisfying both conditions. We emphasize that this conclusion is only valid for exponential coupling constant and for a Ratra-Peebles potential. It would be interesting in a future work to study other type of potential (like the SUGRA model~\cite{brax:2000ys}) and other type of coupling function (like the gaussian coupling constant $A(\phi)=e^{k\phi^2/2}$) which can eventually lead to a different conclusion.

\begin{acknowledgments}
A.~F. thanks A. de Felice for interesting discussions from which this paper originates. A.~H. thanks A. Rivoldini for useful discussions about the statistical analysis.	 A.~Hees is supported by an  FRS-FNRS (Belgian Fund for Scientific Research) Research Fellowship. Numerical simulations were made on the local computing resources (cluster URBM-SysDyn) at the University of Namur (FUNDP).
\end{acknowledgments}

\bibliography{../../../Documents/byMe/biblio}

\end{document}